\documentclass[a4paper,11pt]{article}
\pdfoutput=1 

\usepackage{jheppub} 
\usepackage{amsmath,amssymb,graphicx,slashed,xcolor}
\usepackage{dsfont}
\usepackage{stackengine}
\usepackage{accents}
\usepackage[normalem]{ulem}
\usepackage{soul}

\usepackage[T1]{fontenc} 

\numberwithin{equation}{section}
\newcommand{\nn}{\nonumber \\}

\def\d{{\rm d}}

\graphicspath{{./Figures/}}

\title{\boldmath A new way of calculating the effective potential for a light radion}

\author{J. M.\ Lizana,}
\author{M.\ Olechowski}
\author{and S.\ Pokorski}

\affiliation{Institute of Theoretical Physics, Faculty of Physics,University of Warsaw, Pasteura 5, PL 02-093, Warsaw, Poland}

\emailAdd{javier.lizana@fuw.edu.pl}
\emailAdd{marek.olechowski@fuw.edu.pl}
\emailAdd{stefan.pokorski@fuw.edu.pl}

\abstract{
We address again the old problem of calculating the radion effective potential in Randall-Sundrum scenarios, with the Goldberger-Wise stabilization mechanism. Various prescriptions have been used in the literature, most of them based on heuristic derivations and then applied in some approximations.
We define rigorously a light radion 4D effective action by using the interpolating field method.
For a given choice of the interpolating field, defined as a functional of 5D fields, the radion effective action   is  uniquely defined by the procedure of integrating out the other fields, with the constrained 5D equations of motion always satisfied with help of the Lagrange multipliers. Thus, for a given choice of the interpolating fields  we obtain  a precise prescription for calculating  the effective potential. Different  choices of the interpolating fields give different prescriptions  but in most cases very similar effective potentials. We confirm the correctness of one prescription used so far on a more heuristic basis and also find several new, much more economical, ways of calculating the radion effective potential. Our general considerations are illustrated by several numerical examples. It is shown that in some cases the old methods, especially in models with strong back-reaction, give results which are off even by orders of magnitude. Thus, our results are important e.g.~for estimation of critical temperature in phase transitions.
}

\begin{document} 
\maketitle
\flushbottom

\section{Introduction}
\label{sec:i} 

Warped extra-dimensional models, the original Randall-Sundrum (RS)~model~\cite{Randall:1999ee} and its extensions (see for instance~\cite{Davoudiasl:1999tf, Agashe:2003zs,Contino:2003ve,Falkowski:2008fz,Cabrer:2011fb}), have been proposed as a solution to the hierarchy problem of the Standard Model (SM) -- the quantum instability of the weak scale with respect to the Planck scale. In those scenarios the hierarchy between the Planck scale and the weak scale is generated by the anti-de-Sitter (AdS) warp factor present in the fourth spatial dimension. After integrating out the extra dimension, the four-dimensional spectrum contains brane-localized fields and towers of Kaluza-Klein modes of fields propagating in the bulk. In all such models there is a scalar degree of freedom, the radion, corresponding to  the distance between the branes.

In the original, RS, version the radion remains massless, reflecting the fact that the distance between the branes is not dynamically stabilised. Thus, such model is not able to explain the hierarchy of the Planck and weak scales which may be obtained by arranging the correct separation between the branes. This issue is solved by the Goldberger-Wise mechanism to stabilise the size of the 4-th spatial dimension, by adding a bulk scalar field with some field--dependent potential in the bulk and on the branes~\cite{Goldberger:1999uk}. 
The 4D spectrum of such models contains an infinite KK tower of scalars. The radion is the lightest of them. The stabilisation mechanism generates a mass for the radion but it typically (or at least in large parameter ranges of such models) is much lighter than other 4D states originated from the metric and the bulk scalar~\cite{Goldberger:1999un} (of course excluding the massless graviton).

Various potentials for the Goldberger-Wise scalar have been considered in  the literature~\cite{Goldberger:1999uk,Cabrer:2009we,Bellazzini:2013fga}, categorized according to their back-reaction on the 5D metric, after solving the equations of motion (EOMs) for the scalar-gravity system.

Warped extra dimensional models are very interesting also in other contexts, beyond the  original motivation of solving the hierarchy problem. According to the AdS/CFT correspondence~\cite{Maldacena:1997re}, they provide 5D holographic descriptions of conformal theories in four dimensions, with spontaneously broken conformal symmetry, and perturbed by close-to-mariginal operators. The dynamics of the strongly coupled states in a given 4D theory can be investigated perturbatively by means of the corresponding 5D theory.  In that holographic interpretation, the radion is dual to the dilaton -- the Goldstone boson of the spontaneously broken scale symmetry in 4D theory~\cite{Rattazzi:2000hs, ArkaniHamed:2000ds}.

A useful concept in discussing the stabilisation mechanism and the 4D holographic interpretation of the RS-type models is the effective  radion (dilaton)  potential. It is particularly relevant for investigating  the radion early universe cosmology, such as the character  and the potential impact on the electroweak phase transition of the radion phase transition during which it acquires a vacuum expectation value (corresponding to the dilaton condensation in the holographic picture)~\cite{Creminelli:2001th}.

The radion effective potential has been investigated in several papers.
In some methods~\cite{Luty:2000ec, Bagger:2000eh, Bagger:2003dy, Chacko:2013dra}, an ansatz for the radion as a 5D field is chosen -- the authors assume its composition and profile along the 5-th dimension. Such ansatz is then put in the 5D action and the action is integrated over the 5-th dimension. Some choices of the ansatz can lead to problems like the existence of heavy field tadpoles in the 4D effective theory.
In~\cite{Bellazzini:2013fga, Bunk:2017fic, Megias:2018sxv}, the authors do not introduce any ansatz for the radion, but they restrict to 4D homogeneous configurations to compute the effective potential. Then, when all 5D equations of motion, including boundary conditions at the branes, are fulfilled, the 4D radion has its background value. So this way one obtains only the value of the effective potential for that one configuration. In order to  obtain the radion effective potential,  one has to consider other values of the 4D radion field, that is one has to violate/ignore some of the 5D EOMs. Usually, one ignores the boundary conditions for the warp factor keeping all remaining equations of motion unchanged. We have not found in the literature any precise justification of such a method.

With the goal of having a rigorous definition of the 4D radion effective potential, in this paper we use the interpolating field method (see e.g.~\cite{textbook}). In our case, we choose them for the light 4D states:~the radion and the graviton. We define our interpolating 4D fields as some functionals of all 5D fields (we do not need to choose any anzatz for the 5D properties of the radion; as a result we e.g.~avoid potential problems of heavy tadpoles). For a given choice of the interpolating fields, the radion effective action is  uniquely defined by the procedure of integrating out the other fields, with the constrained 5D EOMs always satisfied with help of the Lagrange multipliers. Thus, for a given choice of the interpolating fields, the result for the effective potential also uniquely follows, with  precise prescription for calculating it. The prescription depends on the choice of the interpolating fields, that is on the functional dependence of the 4D interpolating fields on the original 5D fields. However, for hierarchically heavier all other degrees of freedom and for not very inadequately chosen interpolating fields, although the prescriptions are different the effective potentials themselves are very similar. This can be checked a posteriori, by comparing the results for different interpolating fields, confirming rigorousness of this approach.
We show that the prescription used most often in the literature (see e.g.~\cite{Bellazzini:2013fga, Chacko:2013dra, Bunk:2017fic, Megias:2018sxv}) is correct as it is given by a particular choice of the interpolating fields. However, our  approach leads to alternative prescriptions, equally acceptable, and in particular we find  a new way to calculate the effective potential which is much more efficient in numerical applications.

Obtaining the effective radion potential is a two-step procedure. First, we have to  choose the prescription for calculating it in terms of the original parameters of the 5D model. In the second step, the actual calculation has to be performed. The prescription most often used (again see~\cite{Bellazzini:2013fga, Chacko:2013dra, Bunk:2017fic, Megias:2018sxv}) and confirmed by our approach as a possible choice, requires very high precision numerical calculations. Therefore in the second step, usually, several approximations have been used in  such calculations. Using our new, efficient, method
we have found that in some cases those approximations, e.g.~for the critical temperature  for phase transitions in models with strong back-reaction, give results that are wrong even by orders of magnitude.

In this paper we do not calculate the rescaling factors necessary to normalize the radion fields canonically~\cite{LOP:KT}. However, even non-canonically normalized effective potential is useful for studying the stabilization mechanism and some other aspects of such models. One can identify all extrema of the potential and values of the potential at those extrema and at other characteristic points. The correct calculation of these values is crucial e.g.~for computation of the critical temperature of the radion phase transition.

The paper is organized as follows. Section~\ref{sec:EffPot} is devoted to detailed presentation of our novel approach to calculating the radion effective potential. Some possible choices of the radion interpolating field are also discussed. In Section~\ref{sec:NumComp} we discuss the advantages of the method developed in Section~\ref{sec:EffPot} and use it to calculate explicitly the exact effective potential for several choices of the radion interpolating field. In Section~\ref{sec:Approximate-strong-br} the calculated exact effective potentials are used as a reference for judging about the quality of various approximate calculations  in the literature, based on  the prescription of refs.~\cite{Bellazzini:2013fga, Chacko:2013dra, Bunk:2017fic, Megias:2018sxv}, that we have rigorously justified in Section~\ref{sec:EffPot}. Some more technical issues are presented in four appendices.


\section{Effective action and potential for the radion}
\label{sec:EffPot}

\subsection{Background solution}
\label{sec:5Dmodels} 

We investigate models of gravity and a scalar field $\Phi$ with 5-dimensional (5D) space-time being an orbilofd equal to a warped product of 4-dimensional (4D) space-time and an interval:~${\cal{M}}^5={\cal{M}}^4 \times S^1/{\mathbb Z}_2$. The action involves a bulk potential $V(\Phi)$ and two brane potentials $U_i(\Phi)$ and reads
\begin{equation}
\label{action}
  S
  =
  \frac{1}{2}\int_{{\cal{M}}^4\times S^1}{\d}^5x\left[
  \sqrt{-g}  \left(
  \frac{1}{2\kappa^2}R-\frac12\left(\nabla\Phi\right)^2-V(\Phi)\right)
    -\sum_{i=1,2}\sqrt{-g|_{y_i}}\,\delta(y-y_i)U_i(\Phi)
    \right]\,,
\end{equation}
where $\kappa^2$ is the 5D Einstein's gravitational constant, related to the 5D Planck mass $M_{5D}$ as $\kappa^2=M_{5D}^{-3}$, and $g|_{y_i}$ is the metric in the brane $y=y_i$ inherited from $g$.
We are interested in warped background solutions of the form
\begin{eqnarray}
\d s^2 = g_{MN}\d x^M \d x^N 
&=& e^{-2A(y)}g_{\mu\nu}^{(4)}\d x^\mu \d x^\nu + \d y^2\,,\label{ansatz}
\\
\Phi &=& \phi(y)\,.
\label{ansatz2}
\end{eqnarray}
We will consider two cases for the 4D metric:~the flat Minkowski space-time and inflating, spatially flat de Sitter space-time. Both may be written using  the following 4D metric\footnote{
We use the following convention for the space-time indices: $M,N=0,1,2,3,5$; $\mu,\nu=0,1,2,3$; $i,j=1,2,3$; $x^0=t$; $x^5=y$.}
\begin{equation}
\label{g4}
g_{\mu\nu}^{(4)}\d x^\mu \d x^\nu=-\d t^2 +e^{2Ht}\delta_{ij}\d x^i \d x^j\,,
\end{equation}
where Minkowski case corresponds to $H=0$.
In most of the cases considered in this paper we put $H=0$, but for some analyses it will be useful to keep the possibility of $H\neq 0$ in the equations.

The variations of the action~\eqref{action} for the ansatz~\eqref{ansatz}--\eqref{ansatz2} are explicitly written down in the Appendix~\ref{sec:AcVariation}. Their explicit form will be important below. The variations \eqref{Var1} and \eqref{EoM3b} contain delta-like contributions localized at both ``end-of-the-world'' branes at $y=y_1,y_2$. Thus, it is convenient to write the corresponding equations of motion separately for the bulk (i.e.~for values of $y$ different from the brane positions) and for the branes (obtained by integrating the full equations of motion over an infinitesimally short interval containing a given brane position). The bulk equations read\footnote{
It is convenient to keep their form as they appear in the variations of the action \eqref{Var1}--\eqref{EoM2b}. However, when the equation~\eqref{EoM_extra} is satisfied, it can be used to rewrite~\eqref{EoM_A} in a simpler form as
\begin{equation}
A''=\frac{\kappa^2}{3}(\phi')^2 + e^{2A}H^2\,.
\end{equation}}
\begin{eqnarray}
\label{EoM_phi}
\phi''&=&4A'\phi'+V'\,,
\\
\label{EoM_A}
A^{\prime \prime}&=&2A^{\prime 2}+
\frac{\kappa^2}{6}\phi^{\prime 2}+\frac{\kappa^2}{3}V(\phi)-e^{2A}H^2\,,
\\
\label{EoM_extra}
(A')^2&=&\frac{\kappa^2}{12}(\phi')^2 - \frac{\kappa^2}{6}V+ e^{2A}H^2\,.
\end{eqnarray}
The brane equations of motion are usually written in the form of boundary conditions for derivatives of $\phi$ and $A$:
\begin{eqnarray}
\label{BC_phi}
\lim_{y\to y_i^{\pm}} \phi' &=& \pm \frac12 U'_i(\phi(y_i))\,,
\\
\label{BC_A}
\lim_{y\to y_i^{\pm}} A' &=& \pm \frac{\kappa^2}{6}U_i(\phi(y_i))\,,
\end{eqnarray}
where upper (lower) sign is for $i=1$ ($i=2$) and primes denote derivatives with respect to appropriate argument i.e.~$\d/\d y$ for $A$ and $\phi$ and $\d/\d\phi$ in the case of potentials $V$ and $U_i$.

For general potentials $V$ and $U_i$ the equations of motion \eqref{EoM_phi}--\eqref{EoM_extra} with the boundary conditions \eqref{BC_phi}--\eqref{BC_A} have no solutions. This may be shown by the following reasoning. Let us start at the brane located at $y_1$. There are two dynamical second order equations of motion, \eqref{EoM_phi} and  \eqref{EoM_A}, for two functions:~$\phi(y)$ and $A(y)$. Thus, the initial conditions consist of fixing four values:~$\phi(y_1)$, $\phi'(y_1)$, $A(y_1)$, and $A'(y_1)$. 
The third one corresponds just to the choice of units, so we use the  convention $A(y_1)=0$. The values of the three remaining functions at $y_1$ may be found by solving three equations:~two boundary conditions \eqref{BC_phi} and \eqref{BC_A} for $i=1$ and the bulk equation \eqref{EoM_extra}. Three equations for three unknowns in general may be solved (there may be a discrete set of such solutions). 
Then we may integrate (for example numerically) the dynamical bulk equations of motion, \eqref{EoM_phi} and \eqref{EoM_A}, to find $\phi(y)$ and $A(y)$ in the bulk. The problem appears when we try to fulfill the boundary conditions \eqref{BC_phi} and \eqref{BC_A} at $y=y_2$.\footnote{This does not mean that the necessary fine-tuning is associated with the brane at $y=y_2$.
We could equally well start our integration at brane located at $y_2$ and face problems with
fulfilling the boundary conditions at the brane located at $y_1$.} For general value of $y$ non of these boundary conditions is fulfilled. For some discrete values of $y$ one of them may be fulfilled. 
However, exact fine-tuning of the potential parameters (for example those in $U_2$) is necessary to have solutions to both boundary conditions \eqref{BC_phi} and \eqref{BC_A} at the same value of $y$. This is the standard cosmological constant problem (the effective 4D cosmological constant must vanish for flat 4D sections of the 5D space-time, i.e.~for $H=0$, and must be equal $3H^2$ in the case of inflating 4D sections).

Usually, from the 4D perspective such models are described in terms of KK towers of 4D fields. Such towers are obtained by expanding 5D perturbations around the background \eqref{ansatz}-\eqref{ansatz2} in terms of 4D eigenmodes of the quadratic part of the Lagrangian. The lightest modes of the system are the lightest scalar KK mode (the radion) and a massless spin 2 field (the graviton). 
Reinserting the expansion in the action we could find the interactions among all the 4D fields. 
For some applications it is necessary to go farther and compute the effective action for the radion beyond the perturbative expansion in the number of fields around the vacuum. This is the case, for example, for the study of the cosmological evolution of these models, which includes possible phase transitions in the early universe~\cite{Creminelli:2001th, Randall:2006py, Kaplan:2006yi, Nardini:2007me, Konstandin:2010cd, Bunk:2017fic, Dillon:2017ctw, Megias:2018sxv}, for which non-perturbative solutions interpolating between different vacua are crucial. For such studies it is more convenient to perform an expansion in the number of derivatives in the effective action. In this paper we will focus on the zero-derivative term of this expansion, the effective potential for the radion. Besides its importance for the applications mentioned so far, this object also helps to gain better understanding of the stabilization mechanism of these models~\cite{Goldberger:1999uk,Bellazzini:2013fga,Cox:2014zea}.

\subsection{Definition of the radion effective action}
\label{sec:Definition}

It is known that in order to extract an effective action for low energy degrees of freedom of a model, it is not necessary to know the exact light one-particle mass eigenstates.
In the case of one such light state $|l\rangle$, one can calculate the corresponding effective action using some simple field related to the light degree of freedom in question. The necessary condition is that the state obtained by acting with such a field on the vacuum must not be orthogonal to the light one-particle mass eigenstate:~$\langle l|\Sigma|0\rangle \neq 0$. 
Formally, it is enough that this matrix element is just non-vanishing but in practical applications it is better if it is not very small, as we will discuss later. 
For $n$ light particles $|l_i\rangle$ one needs $n$ fields $\Sigma_j$, which must fulfill the condition that the ($n\times n$)-dimensional matrix $\langle l_i|\Sigma_j|0\rangle$ is invertible:
\begin{equation}
\det\left(\langle l_i|\Sigma_j|0\rangle\right) \neq 0\,. 
\label{InterpolCond} 
\end{equation}

The fields $\Sigma_j$ are usually called in the literature the interpolating fields \cite{textbook}. Once the interpolating fields are chosen, we can integrate out the remaining degrees of freedom to obtain the low energy effective action. Different interpolating fields generate different effective actions, but all of them must be equivalent in the sense of being related by some field redefinitions. 
The well known equivalence theorem for field redefinitions then ensures that all these actions will produce the same S-matrix elements~\cite{Chisholm:1961tha, Kamefuchi:1961sb, Criado:2018sdb}.

The lightest modes of the model under consideration are the graviton and the radion. We assume that the first massive KK graviton is much heavier than the radion (this is really the case for many 5D models and for all models we will use as examples illustrating our general considerations).
Therefore, following the logic explained above, we will define some interpolating 4D graviton and radion fields and integrate out the remaining degrees of freedom.
In realistic models, one expects to have additionally all the Standard Model particles with masses lighter or comparable to the radion mass, but here we will assume that the mixing with them is negligible, so we are only interested in the radion-gravity sector. However, we stress that more general scenarios, where for instance the Higgs-radion mixing is important, could also be treated within the same formalism presented here.

For our system defined by the action \eqref{action} we denote by $\hat h_{\mu\nu}[g_{M,N},\Phi]$ and $\hat\chi[g_{M,N},\Phi]$ such interpolating fields for the graviton and the radion, respectively.
We use the following notation:~$\hat{h}_{\mu\nu}$ and $\hat \chi$ are functionals depending on some arguments (5D metric and 5D scalar) while the same letters without a hat, $h_{\mu\nu}$ and $\chi$, are some specific values of these 4D fields. Of course, functionals $\hat h_{\mu\nu}[g_{M,N},\Phi]$ and $\hat \chi[g_{M,N},\Phi]$ are not arbitrary. They must transform covariantly under 4D diffeomorphisms i.e.~a 5D diffeomorphism applied to the 5D fields, $g_{MN}$ and $\Phi$, should produce a transformation of $\hat h_{\mu\nu}[g_{M,N},\Phi]$ and $\hat \chi[g_{M,N},\Phi]$ given by some 4D diffeomorphism.
Furthermore, $\hat{h}_{\mu\nu}$ must have all the properties that define a 4D Lorentzian metric for all configurations of the 5D fields $g_{MN}$, $\Phi$.
Of course, our interpolating fields should also fulfill the condition \eqref{InterpolCond}.

We calculate the effective action integrating out all the fields except the 4D fields $\hat h_{\mu\nu}$ and $\hat \chi$. This can be expressed as
\begin{equation}
e^{iS_{\mathrm{eff}}[h_{\mu\nu},\chi]}=\int \mathcal{D} g_{MN}\, \mathcal{D} \Phi\, e^{iS} \delta\left(\hat\chi[g_{MN},\Phi]-\chi\right) \delta\left(\hat h_{\mu\nu}[g_{MN},\Phi]-h_{\mu\nu}\right).\label{pathIntegral}
\end{equation} 
Here, the action $S$ inside the path integral is the action~\eqref{action}, and the $\delta$-distributions fix the fields $\hat\chi[g_{MN},\Phi]$ and $\hat h_{\mu\nu}[g_{MN},\Phi]$ to have values equal $\chi$ and $h_{\mu\nu}$, respectively. 
Using the saddle point approximation for the path integral we get:
\begin{align}
S_{\mathrm{eff}}[h_{\mu\nu},\chi]=&\,\textrm{min}~S~\big|_{\mathrm{fixing}~\hat\chi[g_{MN},\Phi]=\chi~\mathrm{and}~\hat h_{\mu\nu}[g_{MN},\Phi]=h_{\mu\nu}}\nn
\equiv& \int \d ^4x \sqrt{-h}\, \mathcal{L}_{\mathrm{eff}}(h_{\mu\nu},\chi).\label{CMinimizationS}
\end{align}
The effective action is therefore a functional of 4D fields $h_{\mu\nu}$ and $\chi$. The minimization in this equation is performed over all possible 5D configurations whose image under the functionals $\hat h_{\mu\nu}$ and $\hat \chi$ is $h_{\mu\nu}$ and $\chi$, respectively.
This constrained minimization can be performed by the Lagrange multiplier method. The equations for the constrained system become
\begin{align}
\frac{\delta S}{\delta \Phi ({\bf x}^{\prime},y)} =&\int \d ^4 {\bf x} \,\left[ \lambda^{\chi}({\bf x}) \frac{\delta \hat \chi({\bf x})}{\delta \Phi({\bf x}^{\prime},y)}+\lambda^{h}_{\mu\nu}({\bf x}) \frac{\delta \hat h^{\mu\nu}({\bf x})}{\delta \Phi({\bf x}^{\prime},y)}\right],\label{EoM1}\\
\frac{\delta S}{\delta g^{MN} ({\bf x}^{\prime},y)} =&\int \d ^4 {\bf x} \,\left[ \lambda^{\chi}({\bf x}) \frac{\delta \hat \chi({\bf x})}{\delta g^{MN}({\bf x}^{\prime},y)}+\lambda^{h}_{\mu\nu}({\bf x}) \frac{\delta \hat h^{\mu\nu}({\bf x})}{\delta g^{MN}({\bf x}^{\prime},y)}\right],\label{EoM2}\\
\hat h_{\mu\nu}[g,\Phi]=&h_{\mu\nu}\,,\label{EoM3}\\
\hat\chi[g,\Phi]=&\chi\,,\label{EoM4}
\end{align} 
where $\lambda^{\chi}({\bf x})$ and $\lambda^{h}_{\mu\nu}({\bf x})$ are the Lagrange multipliers:~new 4D fields whose values have to be determined by solving the system. In the presence of those fields the 5D EOMs are satisfied for the constrained system.

The effective Lagrangian then becomes
\begin{equation}
\sqrt{-h}\, \mathcal{L}_{\mathrm{eff}}(h_{\mu\nu},\chi)=\frac{1}{2}\int_{S^1} \d y \sqrt{-\hat g_{\mathrm{sol}}[h_{\mu\nu},\chi]}\, \mathcal{L}_{\mathrm{5D}}(y,\hat g_{MN\,\mathrm{sol}}[h_{\mu\nu},\chi],\hat \Phi_{\mathrm{sol}}[h_{\mu\nu},\chi]),\label{LEffIntegr5}
\end{equation}
where $\mathcal{L}_{\mathrm{5D}}$ is the 5D Lagrangian of~\eqref{action}, and $\hat g_{MN\,\mathrm{sol}}[h_{\mu\nu},\chi]$, $\hat \Phi_{\mathrm{sol}}[h_{\mu\nu},\chi]$ the solution to the system~\eqref{EoM1}--\eqref{EoM4} given the 4D fields $h_{\mu\nu}$ and $\chi$. Let us stress that we have added hats to $\hat g_{MN\,\mathrm{sol}}$ and $\hat \Phi_{\mathrm{sol}}$ to indicate they are functionals (in this case of 4D fields $h_{\mu\nu}$ and $\chi$), as opposite to $g_{MN}$ and $\Phi$ which are specific 5D field configurations.

It is possible to give a clear meaning to the Lagrange multipliers:~they are the variations of the effective action under the corresponding interpolating fields. This can be checked explicitly. Let $\alpha^i({\bf x})=h_{\mu\nu}({\bf x}),\chi({\bf x})$ be the interpolating fields, and $\omega^n({\bf x},y)=g_{MN}({\bf x},y),\Phi({\bf x},y)$ the 5D fields. Then,
\begin{align}
\frac{\delta S_{\mathrm{eff}}[\alpha]}{\delta \alpha^i({\bf x})}&=\int \d ^4{\bf x}^{\prime}\,\d y\,\frac{\delta S[\hat\omega_{\mathrm{sol}}[\alpha]]}{\delta \omega^n({\bf x^{\prime}},y)} \frac{\delta \hat\omega_{\mathrm{sol}}^n[\alpha]({\bf x}^{\prime},y)}{\delta \alpha^i({\bf x})}\nn
&=\int \d ^4{\bf x}^{\prime}\,\d ^4{\bf x}^{\prime\prime}\,\d y\,
\lambda_j({\bf x}^{\prime \prime}) \frac{\delta \hat \alpha^j[\omega] ({\bf x}^{\prime \prime})}{\delta \omega^n({\bf x^{\prime}},y)} \frac{\delta \hat\omega_{\mathrm{sol}}^n[\alpha]({\bf x}^{\prime},y)}{\delta \alpha^i({\bf x})}\nn
&=\lambda_i({\bf x}),\label{LagMultVar}
\end{align}
where $\lambda_i({\bf x})=\lambda_{\mu\nu}^h({\bf x}),\lambda^{\chi}({\bf x})$ and we have again used the hat notation.
In the first line we have used~\eqref{LEffIntegr5} (integrated over ${\bf x}$), in the second line,~\eqref{EoM2}, and in the third one, the fact that $\hat \alpha^i[\hat \omega_{\mathrm{sol}}[\alpha]]=\alpha$, and therefore,
\begin{equation}
\int \d ^4{\bf x}^{\prime}\, \d y\,\frac{\delta \hat \alpha^j [\omega]({\bf x}^{\prime \prime})}{\delta \omega^n({\bf x^{\prime}},y)} \frac{\delta \hat\omega_{\mathrm{sol}}^n[\alpha]({\bf x}^{\prime},y)}{\delta \alpha^i({\bf x})}=\delta^j_i\,\delta({\bf x}^{\prime \prime}-{\bf x}).
\end{equation}
Relation \eqref{LagMultVar} will prove very useful for obtaining the effective action for the radion.

In Appendix~\ref{sec:AnsatzMethod} we review the ansatz method used in~\cite{Luty:2000ec, Bagger:2000eh, Bagger:2003dy, Chacko:2013dra} and show that the same results may be obtained more directly, and without heavy field tadpoles, with the interpolating field method used in this paper. The interpolating field method has advantages, when compared to the ansatz method, which we discuss later.

\subsection{Expansion of the effective action}
\label{sec:DerivativeExp}

For a given choice of the interpolating fields, using the diffeomorphism invariance, we can write the effective Lagrangian using the expansion 
\begin{align}
\mathcal{L}_{\mathrm{eff}}(h_{\mu\nu},\chi)
= -V_{\mathrm{eff}}(\chi)-\frac{1}{2}\,C_{\mathrm{eff}}(\chi)h^{\mu\nu}\partial_{\mu} \chi \partial_{\nu} \chi +\frac{1}{2}K_{\mathrm{eff}}(\chi) R[h_{\mu\nu}]+\dots,\label{EffActV}
\end{align}
where the ellipsis represents terms of higher order in derivatives and/or in the curvature. This expansion defines the effective potential $V_{\mathrm{eff}}(\chi)$, the kinetic term $C_{\mathrm{eff}}(\chi)$, and the kinetic mixing with gravity $K_{\mathrm{eff}}(\chi)$ for the radion.


The effective potential $V_{\mathrm{eff}}(\chi)$ can be calculated computing the effective Lagrangian for the 4D flat metric $h_{\mu\nu}\propto\eta_{\mu\nu}$ and homogeneous configurations of $\chi({\bf x})=\chi$. Then, the terms involving curvature tensors and derivatives vanish and we have $V_{\mathrm{eff}}(\chi)=-\mathcal{L}_{\mathrm{eff}}(\eta_{\mu\nu},\chi)$. In this case, the equations to find the 5D fields $\hat g_{MN\,\mathrm{sol}}$ and $\hat \Phi_{\mathrm{sol}}$ are Minkowski invariant, so we set $H=0$ in the ansatz~\eqref{ansatz}--\eqref{g4}. The Lagrange multipliers obtained with these configurations can be related to $V_{\mathrm{eff}}(\chi)$ through~\eqref{LagMultVar} because higher derivative terms in $\chi$ and $h_{\mu\nu}$ vanish. Using 4D Lorentz covariance to rewrite the Lagrange multiplier $\lambda^{h}_{\mu\nu}$ as $\lambda^{h}_{\mu\nu}=\lambda^{h} h_{\mu\nu}$, we obtain
\begin{align}
\lambda^h
&=\frac{1}{2}\sqrt{-h}\, V_{\mathrm{eff}}(\chi),\label{Vthh}\\
\lambda^{\chi}&=-\sqrt{-h}\,\frac{\d }{\d \chi} V_{\mathrm{eff}}(\chi).\label{Vthchi}
\end{align}
These formulae provide a method to compute the effective potential rigorously defined for a particular choice of the interpolating fields. In particular, as we show later, for one specific choice of the interpolating fields \eqref{Vthh} gives the prescription for computing the effective potential for soft wall models that has been most often used in the literature~\cite{Bellazzini:2013fga,Bunk:2017fic,Megias:2018sxv}. We also show that different choices of interpolating fields give different prescriptions  based on the same~\eqref{Vthh}, which could be more convenient in some situations. However,~\eqref{Vthchi}  provides a new method for the same computation which is much more economical in numeric calculations.

Some attention should be given to the radion-graviton mixing present in ~\eqref{EffActV}. 
Such kinetic mixing with gravity can be eliminated through a Weyl transformation of the metric. We can always go to the Einstein frame by doing the transformation
\begin{equation}
\tilde h_{\mu\nu}= \frac{K_{\mathrm{eff}}(\chi)}{K_{\mathrm{eff}}(\chi^{*})}\, h_{\mu \nu},
\end{equation}
where $\chi^{*}$ is some reference value of the radion. The effective Lagrangian is then
\begin{align}
\tilde {\mathcal{L}}_{\mathrm{eff}}(\tilde h_{\mu\nu},\chi)
= -\tilde V_{\mathrm{eff}}(\chi)-\frac{1}{2}\,\tilde C_{\mathrm{eff}}(\chi)\tilde h^{\mu\nu}\partial_{\mu} \chi \partial_{\nu} \chi +\frac{M_P^2}{2} R[\tilde h_{\mu\nu}]+\dots,\label{EffActVC}
\end{align}
where
\begin{align}
\tilde V_{\mathrm{eff}}(\chi)=& \frac{K_{\mathrm{eff}}^{2}(\chi^*)}{K_{\mathrm{eff}}^{2}(\chi)} V_{\mathrm{eff}}(\chi),\label{CanVeff}\\
\tilde C_{\mathrm{eff}}(\chi)=&\frac{K_{\mathrm{eff}}(\chi^*)}{K_{\mathrm{eff}}(\chi)}C_{\mathrm{eff}}(\chi)+\frac{3}{2}K_{\mathrm{eff}}(\chi^*)\frac{K^{\prime\,2}_{\mathrm{eff}}(\chi)}{K_{\mathrm{eff}}^2(\chi)},
\end{align}
and the 4D Planck mass is given by $M_P^2=K_{\mathrm{eff}}(\chi^*)$.{\footnote{In some cases the simplest possibility is to take the radion vacuum value as the reference point, $\chi^*$, but this is not necessary.}}
The calculation of $K_{\mathrm{eff}}$ is therefore necessary to translate natural 5D units of our system to physical units by matching the value of the 4D Planck mass. In addition, $K_{\mathrm{eff}}$ rescales also the kinetic term $C_{\mathrm{eff}}$ and the effective potential $V_{\mathrm{eff}}$.\footnote{In general, vacuum i.e.~the configuration which minimizes the rescaled potential $\tilde V_{\mathrm{eff}}(\chi)$ is different from that minimizing $V_{\mathrm{eff}}$. However, in the most interesting case that the vacuum solution has flat 4D sections, both minima of $\tilde V_{\mathrm{eff}}$ and $V_{\mathrm{eff}}$ coincide because both are located at the point $\chi_{\mathrm{min}}$ where $\tilde V_{\mathrm{eff}}(\chi_{\mathrm{min}})=V_{\mathrm{eff}}(\chi_{\mathrm{min}})=0$.}
It is possible to obtain the expression for the kinetic mixing with gravity, $K_{\mathrm{eff}}$, using the value of the graviton interpolating field $h_{\mu\nu} \propto g^{(4)}_{\mu\nu}$ of~\eqref{g4} with $H\neq 0$, instead of~\eqref{Vthh}. The derivation is presented in the Appendix~\ref{sec:KineticMixingG} and here we just give the result:
\begin{equation}
K_{\mathrm{eff}}(\chi)= \frac{1}{\kappa^2}  \int_{y_1}^{y_2} \d y\, e^{-2 A_{\mathrm{sol}}}, \label{KForm}
\end{equation}
where $A_{\mathrm{sol}}$ is the solution for given homogeneous $\chi$ and $h_{\mu\nu}=\eta_{\mu\nu}$.

To properly determine the dynamics of the radion, in particular  its mass, we also need to compute the kinetic term $C_{\mathrm{eff}}(\chi)$ and the higher orders in~\eqref{EffActV}.
This work is however mainly devoted to the effective potential and its calculation. A thorough analysis of the kinetic term will be presented in a follow-up article. In any case, the effective potential in the Einstein frame already provides valuable physical information about the system, e.g.~the energy difference between different extrema of the potential. Such quantities are fundamental for the computation of critical temperatures at which phase transitions may take place~\cite{Bunk:2017fic,Megias:2018sxv}.

It remains to discuss various aspects of the dependence of the effective potential on the choice of the interpolating fields. 
If no truncation of the derivative and curvature expansion \eqref{EffActV} was performed, all the corresponding effective actions are equivalent (related by some field redefinitions) and give the same physics, unless the overlap of some of the interpolating fields with the corresponding light mass eigenstates (in our case: the graviton and the radion) is exactly zero.
The quality of approximation based on a truncated~\eqref{EffActV} depends on two factors. When the interpolating fields are chosen as the canonically normalized light mass eigenstates, the terms with more than two derivatives are suppressed by appropriate powers of masses of the integrated out heavy states. The effective potential is then defined unambiguously up to such corrections. In particular, its second derivative gives the physical mass.
However, in the case of choosing different interpolating fields there appear in addition differences in the effective Lagrangians truncated to the same operators. These remarks apply to the effective potentials in the Einstein frame calculated for different {\it canonically normalized} interpolating fields. The difference between them is compensated by higher derivative and curvature terms that are suppressed by heavy masses, which can however  be large (with large coefficients) if an interpolating field is close to being orthogonal to the mass eigenstate one. The question is how one can judge if the choice of the interpolating field is good or bad, in the  sense of giving the effective potential {\it after canonical normalization} close to the one obtained for mass eigenstates. A similarity of the effective potentials for different {\it canonically normalized} interpolating fields would be a strong hint for their proper choice (note that different good choices can produce very different kinetic and mixing terms). Since we do not calculate the canonical normalization of the interpolating fields in this paper we cannot perform such a direct comparison.  We discuss this issue in the next section. Of course,
another test of good or bad choices is to check their overlap with the mass eigenstates.


\subsection{Interpolating fields}
\label{sec:InterFields}

To compute the effective potential we must specify the interpolating fields to be used. In this subsection we will list some possibilities and discuss them.

\subsubsection{The UV metric and the IR warp factor as interpolating fields}

The prescription to calculate the radion effective potential most often used in the literature~\cite{Bellazzini:2013fga,Bunk:2017fic,Megias:2018sxv} may be justified by our procedure with the following interpolating fields for the  metric and the radion
\begin{align}
\hat h_{\mu\nu}[g,\Phi]&=\left.g_{\mu\nu}\right|_{y_1}\,,\label{defh1}\\
\hat \chi_g[g,\Phi]&=(-g|_{y_2})^{1/8}(-g_{y_1})^{-1/8}
=\exp\left[-\left(A|_{y_2}-A|_{y_1}\right)\right],\label{defchi1}
\end{align}
where $g|_{y_i}$ is the metric in the brane $y=y_i$ inherited from $g$, and the ansatz~\eqref{ansatz} was used in the last line.
We have added the subscript $g$ to the interpolating radion field $\chi_g$ to distinguish it from other choices. The advantage of the above choice is its simplicity. Both interpolating radion and graviton fields depend only on the fields evaluated at the branes. The fact that we have taken as interpolating graviton field the metric on the UV brane, and not on the IR brane, has to do with the resulting $K_{\mathrm{ef f}}$ function. Choice \eqref{defh1} leads to a very weak dependence of $K_{\mathrm{eff}}$ on $\chi$ as we will see below.

For the interpolating fields \eqref{defh1} and \eqref{defchi1}, the Lagrange multipliers introduce modifications of EOMs~\eqref{EoM_phi}--\eqref{BC_A} due to the r.h.s.~of~\eqref{EoM1} and~\eqref{EoM2}. The non-vanishing variations of the interpolating fields are:
\begin{align}
\frac{\delta \hat h^{\rho\sigma} ({\bf x})}{\delta g^{\mu \nu}({\bf x}^{\prime},y)} =&\delta^{\rho}_{(\mu} \delta^{\sigma}_{\nu)} \,\delta({\bf x}-{\bf x}^{\prime})\,\delta(y-y_1),\label{Varhg}\\
\frac{\delta \hat \chi_g ({\bf x})}{\delta g^{\mu \nu}({\bf x}^{\prime},y)} =&\frac{\chi_g({\bf x})}{8} \delta({\bf x}-{\bf x}^{\prime})\left(\delta(y-y_1) g_{\mu\nu}|_{y_1}-\delta(y-y_2)g_{\mu\nu}|_{y_2}\right).\label{VarChig}
\end{align}
Therefore, using~\eqref{EoM2} and~\eqref{EoM3b}, one can see that the UV and IR Israel junction conditions~\eqref{BC_A} are replaced with
\begin{align}
 \left.e^{-4A}\left[U_{1}(\phi)-\frac{6}{\kappa^2}A^{\prime}\right]\right|_{y_1^+} =&\frac{\lambda^{\chi}\chi_g}{2}  +4\lambda^h,\label{BCA1}\\
\left.e^{-4A}\left[ U_{2}(\phi)+\frac{6}{\kappa^2}A^{\prime}\right]\right|_{y_2^-}=&-\frac{\lambda^{\chi}\chi_g}{2}.\label{BCA2}
\end{align}
The Lagrange multipliers break the Israel junction conditions, and so  they allow to find solutions to the EOMs~\eqref{EoM1} and~\eqref{EoM2} for given values of the interpolation graviton and radion fields~\eqref{EoM3}.
Following~\eqref{Vthh}, the effective potential is proportional to $\lambda^h$. It follows from~\eqref{BCA1} and~\eqref{BCA2} that this effective potential may be written as a sum of two terms:
\begin{equation}
V_{\mathrm{eff}} = V^{\mathrm{UV}}_{\mathrm{eff}}+V^{\mathrm{IR}}_{\mathrm{eff}}\,,\label{EfPot1} 
\end{equation}
where
\begin{eqnarray}
V^{\mathrm{UV}}_{\mathrm{eff}} &=&\frac{1}{2}\left.\left[-\frac{6}{\kappa^2}A^{\prime}+U_1(\phi)\right]\right|_{y_1^+},\label{VeffUV}\\
V^{\mathrm{IR}}_{\mathrm{eff}} &=&\frac{1}{2}e^{-4(A(y_2)-A(y_1))}\left.\left[\frac{6}{\kappa^2}A^{\prime}+U_2(\phi)\right]\right|_{y_2^-}.\label{VeffIR}
\end{eqnarray}
The first of the above terms is exclusively evaluated at the UV brane. The second involves quantities evaluated at both branes. It looks like being exclusively evaluated at the IR brane if convention is used in which $A(y_1)$ is fixed (we use a common convention in which $A(y_1)=0$). In the literature these terms are usually called, respectively,  the ultraviolet and the infrared contribution to the effective potential~\cite{Bellazzini:2013fga}. We use similar notation, $V^{\mathrm{UV}}_{\mathrm{eff}}$ and $V^{\mathrm{IR}}_{\mathrm{eff}}$, but we want to stress that these terms do not represent contributions to the radion potential physically localized at the branes. Decomposition of $V_{\mathrm{eff}}$ into such two terms is just a consequence of using formula \eqref{Vthh} and the interpolating fields \eqref{defh1} and \eqref{defchi1}.\footnote{Even defining the effective potential as the integral of the 5D potential over the 5-th dimension, as done e.g~in \citep{Bellazzini:2013fga}, one gets contributions to $V_{\mathrm{eff}}$ from both branes and from the bulk. The result may be written as a difference of two boundary terms if integration by parts is applied. However, integration by parts is just a tool to calculate integrals in which ``contribution'' from a given boundary is known only up to an arbitrary constant.} As we will show later, other decompositions appear for other methods of finding $V_{\mathrm{eff}}$ and/or other choices of the interpolating fields.

Thus, we rigorously confirm the prescription used e.g.~in \cite{Bellazzini:2013fga,Bunk:2017fic,Megias:2018sxv} for calculating the effective potential for the particular choice of the interpolating fields. In the next step, in the actual calculation of the radion effective potential from eq.~\eqref{EfPot1}, the same definition of the interpolating fields has to be consistently used. As we shall discuss it in Section~\ref{sec:NumComp}, such calculations require extremely high precision and various approximations have been used in the literature, including an inconsistent use of radion (interpolating) fields. In Section~\ref{sec:Approximate-strong-br} we critically review those approximations and compare them  with our rigorous results.

Before we discuss other choices of interpolating fields we discuss the method based on~\eqref{Vthchi}.
As we already mentioned, relation~\eqref{Vthchi} provides a new way of calculating the effective potential using~$\lambda^{\chi}$. Equation~\eqref{Vthchi} determines $V_{\mathrm{eff}}$ up to some overall shift.  This shift may be easily obtained from the condition that the effective potential vanishes for the radion corresponding exactly to the background position of the brane - for which all EOMs and BCs are fulfilled. 
For the choice of the interpolating fields given in \eqref{defh1}--\eqref{defchi1}, $\lambda^{\chi}$ can be calculated from the breaking of the IR Israel junction condition~\eqref{BCA2}:
\begin{align}
\frac{\d V_{\mathrm{eff}}(\chi_g)}{\d \chi_g}=\left.2e^{-4A}\chi_g^{-1}\left[\frac{6}{\kappa^2}A^{\prime}+U_2(\phi)\right]\right|_{y=y_2^-}
=4\chi_g^{-1}V^{\mathrm{IR}}_{\mathrm{eff}}(\chi_g).\label{VIRdV}
\end{align}
We stress that this relation between the derivative of the full effective potential and the IR part had not been appreciated before, and it constitutes one of the main results of this article.\footnote{Similar  formula, used in a more specific context, is also given in~\cite{Pomarol:2019aae}. We thank the authors of~\cite{Pomarol:2019aae} for drawing our attention to it.}
It provides a much more efficient (and much faster) method of numerical computation of the effective potential.

The choice \eqref{defh1} for the graviton interpolating field has one more advantage (in addition to its simplicity). For the systems we are interested in (and which will be discussed in more detail in the next section) it leads to the kinetic mixing between radion and gravity, described by $K_{\mathrm{eff}}(\chi)$, which depends very weakly on $\chi$. As explained in Appendix \ref{sec:KineticMixingG}, $K_{\mathrm{eff}}(\chi)$ may be in such situation approximated by an easy to calculate constant. Thus, we are able to determine with good accuracy the value of the 4D Planck mass predicted by a given model.

\subsubsection{Other interpolating fields}
\label{sec:OtherInterFilds}

One can also use interpolating fields $\hat h_{\mu\nu}[g,\Phi]$ and $\hat \chi[g,\Phi]$ different than \eqref{defh1}-\eqref{defchi1}. 
Let us first change our choice for the graviton interpolating field while keeping \eqref{defchi1} for the radion. We choose the metric evaluated at the infrared brane:
\begin{equation}
\hat h^{\mathrm{IR}}_{\mu\nu}[g,\Phi]=\left.g_{\mu\nu}\right|_{y_2}. \label{defhIR}
\end{equation}
Then, the application of the methods explained above gives the effective potential which may be expressed in terms of \eqref{EfPot1}-\eqref{VeffIR} as:
\begin{equation}
\bar V_{\mathrm{eff}}(\chi)=\chi^{-4} V_{\mathrm{eff}}(\chi)\,.
\label{Veffbar}
\end{equation}
(We use bars to denote quantities obtained with~\eqref{defhIR}, as opposite to the unbarred ones, obtained with~\eqref{defh1}.)
The radion-gravity mixing is also related in both cases
\begin{equation}
\bar K_{\mathrm{eff}}(\chi)= \chi^{-2}K_{\mathrm{eff}}(\chi)\,.
\label{Keffbar}
\end{equation}
Thus, although $\bar V_{\mathrm{eff}}$ differs from 
$V_{\mathrm{eff}}$, the effective potential in the Einstein frame is exactly the same for both choices. This is somewhat peculiar for this choice as a  consequence of the fact that with both interpolating graviton fields we are breaking the same equations of motion. In general, for other radion interpolating fields we expect slightly different effective potentials in the Einstein frame when we change the graviton interpolating field.

Using \eqref{Vthchi} one can find (in analogy to \eqref{VIRdV}) the relation between the derivative of the effective potential and one of its parts\footnote{This potential may be written as a sum of two terms,
$\bar V_{\mathrm{eff}}(\chi)=\bar V_{\mathrm{eff}}^{\rm UV}(\chi)+\bar V_{\mathrm{eff}}^{\rm IR}(\chi)$,
with $\bar{V}_{\mathrm{eff}}^{\rm IR}(\chi)$ determined by quantities evaluated only at the IR brane while $\bar{V}_{\mathrm{eff}}^{\rm UV}(\chi)$ depends on quantities evaluated at both branes. This is analogous to the previous case just with the roles of branes interchanged.
}
\begin{equation}
\chi\frac{\d  \bar V_{\mathrm{eff}}(\chi)}{\d  \chi}
=-4\bar{V}^{\mathrm{UV}}_{\mathrm{eff}}
=-4\chi^{-4}V_{\mathrm{eff}}^{\mathrm{UV}}(\chi)\label{VUVdV}\,.
\end{equation}
Notice that now the derivative of the potential is related to the UV part of the potential while before it was related to the IR part.

Let us now take again~\eqref{defh1} as the graviton interpolating field  but explore other choices for the radion interpolating field.
Some possibilities are:~the physical distance between branes
\begin{equation}
\hat \chi_y[g,\Phi]=\frac{1}{2}\int_{S^1} \d y \sqrt{g_{55}}\,,\label{chidistance}
\end{equation}
and the value of the field $\phi$ on the IR brane\footnote{In the limit of vanishing back-reaction, the condition~\eqref{InterpolCond} is no longer satisfied for $\hat\chi_{\phi}$, so the effective action (and therefore the effective potential) fails to provide a good description of light degrees of freedom. However, for models with strong back-reaction in the IR, $\hat\chi_{\phi}$ could  in principle be an acceptable option. We will see in Section~\ref{sec:NumComp} that in any case, it is a poor choice.}
\begin{equation}
\hat\chi_{\phi}=\Phi(y_2)\,.
\label{chiPhiy2}
\end{equation}
Non-vanishing variations of these interpolating fields, for our ansatz \eqref{ansatz}-\eqref{ansatz2}, are
\begin{align}
\frac{\delta \hat \chi_y ({\bf x})}{\delta g^{55}({\bf x}^{\prime},y)} =&
-\frac{1}{4}\delta({\bf x}-{\bf x}^{\prime}),\\
\frac{\delta \hat \chi_{\phi} ({\bf x})}{\delta \Phi({\bf x}^{\prime},y)} =&
\delta({\bf x}-{\bf x}^{\prime})\,\delta(y-y_2).
\end{align}
For both choices, $\lambda^h$ gives the only contribution to the UV Israel junction condition breaking:
\begin{equation}
\left.e^{-4A}\left[U_{1}(\phi)-\frac{6}{\kappa^2}A^{\prime}\right]\right|_{y_1^+} =4\lambda^h,\label{BCA1B}\\
\end{equation}
leading to the effective potential with only the UV contribution:
\begin{equation}
V_{\mathrm{eff}}^{(y)} =V_{\mathrm{eff}}^{(\phi)} =\frac{1}{2}\left.\left[-\frac{6}{\kappa^2}A^{\prime}+U_1(\phi)\right]\right|_{y=y_1^+}.\label{EfPot2}
\end{equation}

We use the superscript $(y)$ or $(\phi)$ to indicate that the effective potential was obtained using $\hat \chi_y$ or $\hat \chi_{\phi}$, respectively, as the interpolating radion field.
In the case of $\hat \chi_y$, the Lagrange multiplier $\lambda^{\chi}$ modifies the constraint equation~\eqref{EoM_extra} to
\begin{equation}
e^{-4A(y)}\left[\frac{6}{\kappa^2}A^{\prime 2}(y)-\frac{\phi^{\prime 2}(y)}{2} +V(\phi(y))\right]=-\lambda^{\chi}.
\label{ConstrMod}
\end{equation}
Any value of $y$ may be used in the above equation because the dependence of the l.h.s.~on $y$ vanishes if the remaining bulk EOMs are satisfied. Equation~\eqref{Vthh} gives different expression for the effective potential which in the present case reads
\begin{equation}
\frac{\d V^{(y)}_{\mathrm{eff}}(\chi_y)}{\d \chi_y}=e^{-4(A(y)-A(y_1))}\left[\frac{6}{\kappa^2}A^{\prime 2}(y)-\frac{\phi^{\prime 2}(y)}{2} +V(\phi(y))\right],\label{dVeffPot2}
\end{equation}
and the r.h.s.~may be evaluated at any value of $y$.

Using $\hat \chi_{\phi}$ as the interpolating field, 
we find the following modification of the boundary condition 
\eqref{BC_phi} at the IR brane
\begin{equation}
\left.e^{-4A}\left[\phi^{\prime}+\frac{1}{2}U^{\prime}_2(\phi)\right] \right|_{y=y_2^{-}}=-\lambda^{\chi},
\end{equation}
Thus, the derivative of the effective potential may be written as
\begin{equation}
 \frac{\d V^{(\phi)}_{\mathrm{eff}}(\chi_{\phi})}{\d \chi_{\phi}}=\left.e^{-4(A(y)-A(y_1))}\left[\phi^{\prime}+\frac{1}{2}U^{\prime}_2(\phi)  \right]\right|_{y=y_2^{-}}.\label{dVeffPot3}
\end{equation}

In this section we have derived several prescriptions, based on  eqs.~\eqref{Vthh}  and~\eqref{Vthchi},  
 to compute the radion effective potential for different interpolating fields. Of course, the number of possible interpolating fields is infinite. Here we have just reviewed some simple choices. In the next section we use them for the actual numerical computation.

\section{Computation of the radion effective potential}
\label{sec:NumComp}

The purpose of this section is two-fold. First, we demonstrate the superiority of the  approach based on~\eqref{Vthchi} over the one using~\eqref{Vthh}. Secondly, we investigate the stability of the radion effective potential  against changes of the interpolating fields, that is changes of the prescriptions  to calculate it.

We consider two different types of bulk potentials for the scalar bulk field $\Phi$. First, a quadratic potential (used for example in~\cite{Bellazzini:2013fga,Bunk:2017fic,AbuAjamieh:2017khi}):
\begin{equation}
V_{\mathrm{Quad}}(\Phi)=-\frac{6k^2}{\kappa^2}-2\epsilon k^2\Phi^2.\label{QuadBulkV}
\end{equation}
This is a typical potential used in the Goldberger-Wise mechanism~\cite{Goldberger:1999uk}. In order to obtain a large hierarchy between the Planck and TeV scales in a natural way, one usually considers small values of $|\epsilon|$ ($\leq 0.1$). In this paper we will additionally focus on positive $\epsilon$, which typically results in asymptotically AdS spaces with strong back-reaction only close to the IR brane.
We also consider the following exponential bulk potential (used in~\cite{Megias:2018sxv,Cabrer:2009we}):\footnote{The apparently capricious form of the exponential potential follows from the simple superpotential
\begin{equation}
W(\Phi)=\frac{6k}{\kappa^2}\left( 1+e^{\gamma \Phi} \right).
\end{equation}}
\begin{equation}
V_{\mathrm{Exp}}(\Phi)=-\frac{6k^2}{\kappa^2}\left[1+2e^{\gamma \Phi}+\left(1-\frac{3 \gamma^2}{4 \kappa^2}\right)e^{2 \gamma \Phi}  \right],\label{ExpBulkV}
\end{equation}
with the assumption $3\gamma^2<4\kappa^2$ for which the potential is monotonically decreasing.\footnote{This potentials have been also used for models that substitute the IR brane for a singular behaviour of the metric and the scalar field. For these models, this assumption is additionally fundamental because of consistency reasons~\cite{Cabrer:2009we}.}

Both potentials have in common that they are bounded from above by $-6k^2/\kappa^2$ but are not bounded from below. 
This in principle is not problematic for $(3+1)$--D Poincar\'e invariant solutions of five-dimensional gravity coupled to scalars~\cite{Gubser:2000nd}. Besides, in asymptotically AdS spaces, unitarity bounds allow for negative masses squared as long as the Breitenlohner-Freedman bound is satisfied (for the potential~\eqref{QuadBulkV} this leads to the bound $\epsilon<1$)~\cite{Breitenlohner:1982jf}.

\begin{figure}[t!]
\begin{center}
\includegraphics[height=5.3cm]{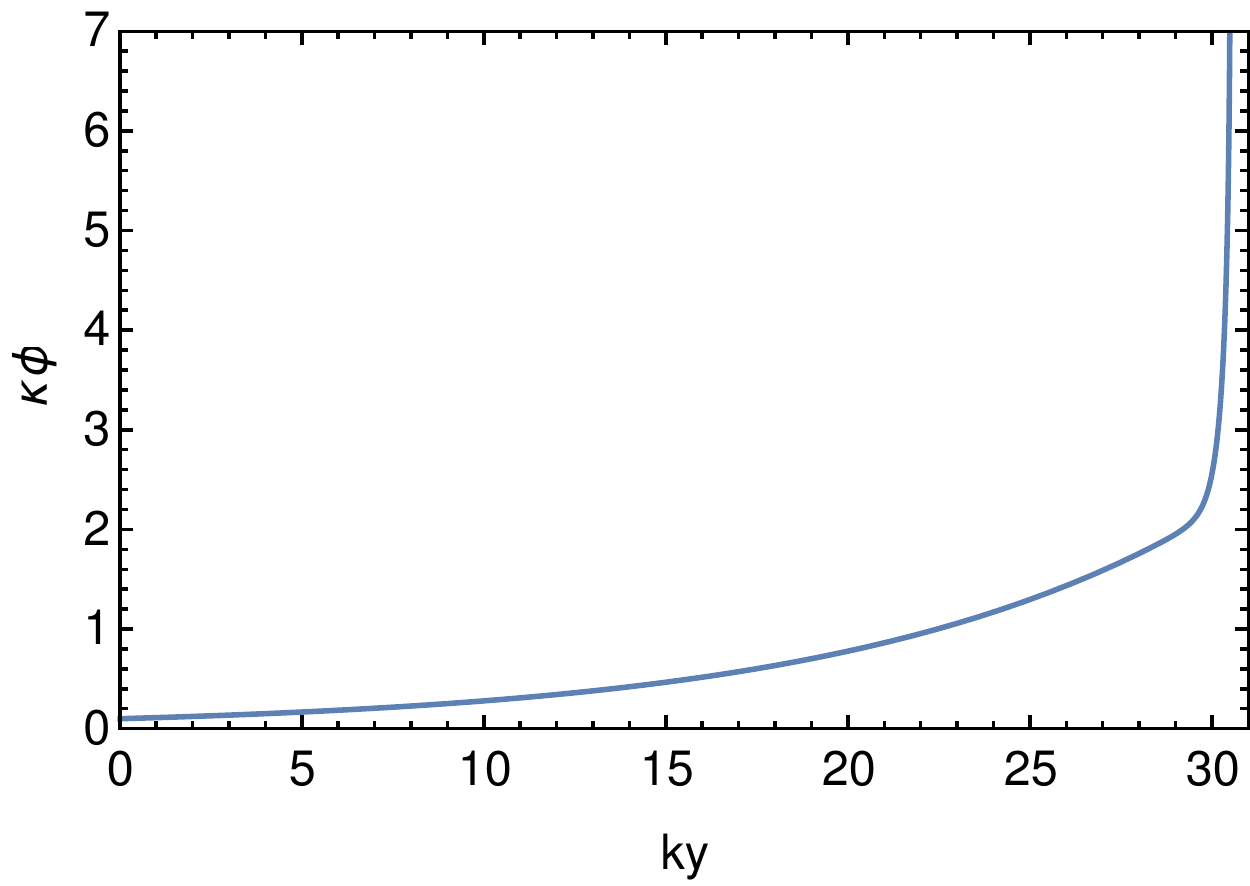}
\includegraphics[height=5.3cm]{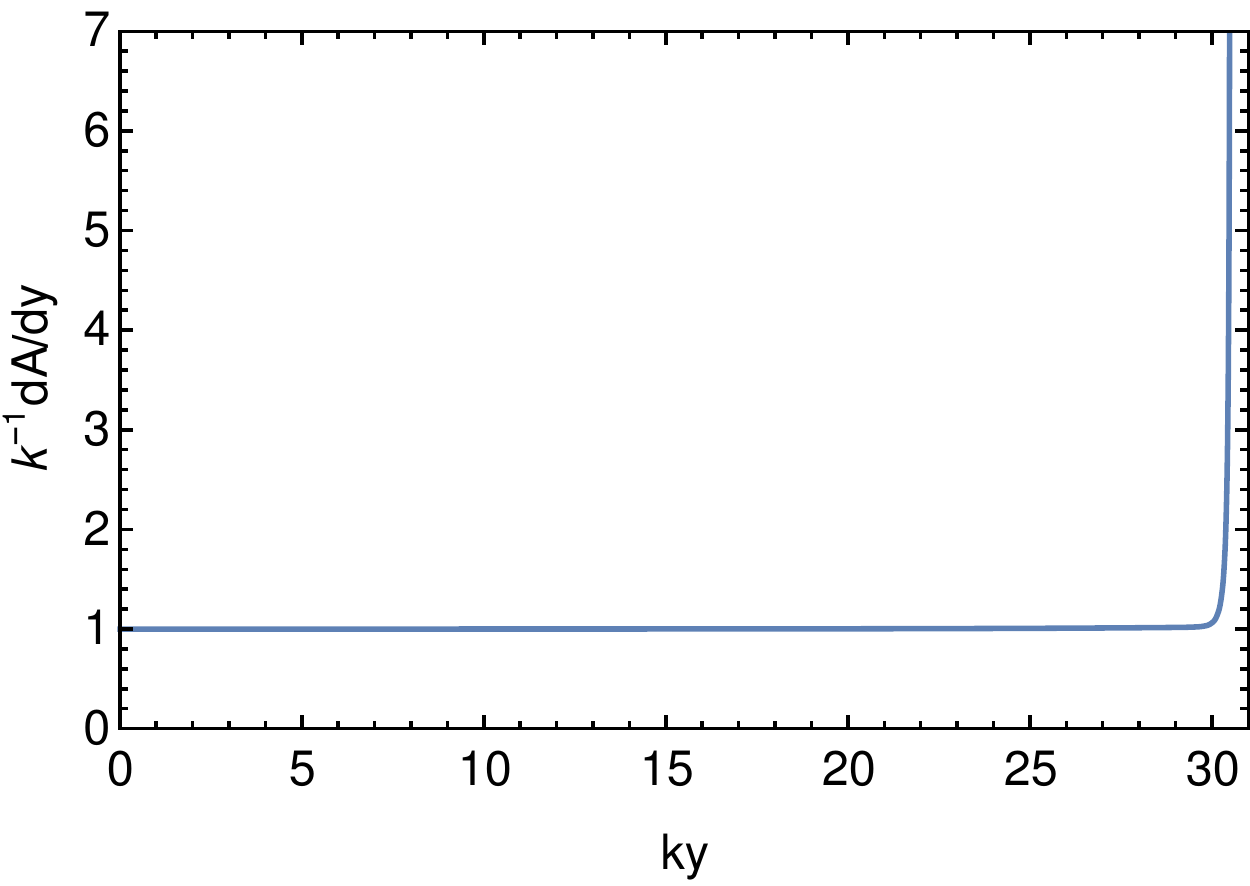}
\caption{Typical background solution to the bulk equations~\eqref{EoM_phi}-\eqref{EoM_extra}. For this case, we have taken the quadratic potential with $\kappa=0.5\,k^{-3/2}$ and $\epsilon=0.1$.}
\label{fig:BackGround}
\end{center}
\end{figure}

For the brane potentials we consider quadratic ones:
\begin{equation}
U_{1,2}(\Phi)=\Lambda_{1,2}+\xi_{1,2}(\Phi-v_{1,2})^2,\label{branePotForm}
\end{equation}
where $\Lambda_{1,2}$ are the tensions of the UV and IR branes, respectively. In most of the cases we will take the so-called stiff wall approximation, where
$\xi_{1,2}\to\infty.$ In such a case, the boundary conditions for 
$\phi$~\eqref{BC_phi} and $A$~\eqref{BC_A} simplify to
\begin{align}
\lim_{y\to y_i^{\pm}}  \phi(y)&=v_{i}\,,\label{StiffBC_phi}\\
\lim_{y\to y_i^{\pm}}  A^{\prime}(y)&=\pm\Lambda_{i}\,,\label{StiffBC_A}
\end{align}
where the plus (minus) sign is taken for the subscript $i=1$ ($i=2$). Also, if $\phi(y)$ satisfies the boundary condition~\eqref{BC_phi},
\begin{equation}
\lim_{\xi_i\to\infty}U_{i}(\phi(y_i))\to\Lambda_i.\label{StiffU}
\end{equation}
In Figure~\ref{fig:BackGround} we show the solution of the EOMs~\eqref{EoM_phi}--\eqref{BC_A} for the quadratic potential with $\kappa=0.5\,k^{-3/2}$ and $\epsilon=0.1$. Similar background solutions are typical for both potentials \eqref{QuadBulkV} and \eqref{ExpBulkV}.
One can see that in the UV region (small values of $y$), the metric is very close to the AdS case (for which $A^\prime=\mathrm{const}$), but close to the IR brane, large deviations from the AdS metric appear. These are the so-called soft wall models, which have been vastly studied in detail due to their exceptional suitability for constructing phenomenologically interesting models~\cite{Falkowski:2008fz, Cabrer:2011fb, Carmona:2011ib, Cabrer:2011qb, Megias:2017ove, Megias:2019vdb}.

\subsection{Warp factor as interpolating radion field}

From now on we use the interpolating graviton field given by~\eqref{defh1}. For the interpolating radion field, we focus first on $\hat\chi_g$ defind in~\eqref{defchi1}.
The effective potential for this case is given by~\eqref{EfPot1}. In order to evaluate this formula one has to find a class of solutions to the EOMs~\eqref{EoM_phi}-\eqref{BC_phi} which leaves one free degree of freedom (e.g.~the distance between the branes). Then, \eqref{EfPot1} may be used to compute the effective potential as a function of this degree of freedom. In addition, it is necessary to find the value of the radion field $\chi_g$ as a function of the same degree of freedom. Combining these results one can write the obtained effective potential as a function of the radion.
However, in general it is not possible to perform this procedure analytically so the use of numerical computations is necessary.

\begin{figure}[t!]
\begin{center}
\includegraphics[height=5.3cm]{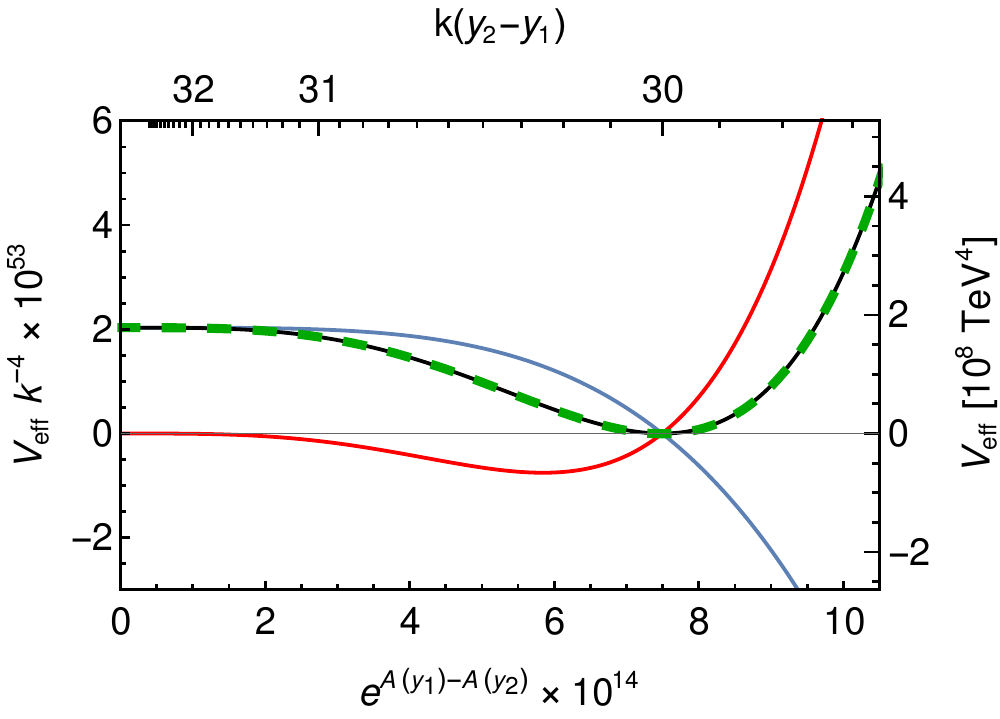}
\includegraphics[height=5.3cm]{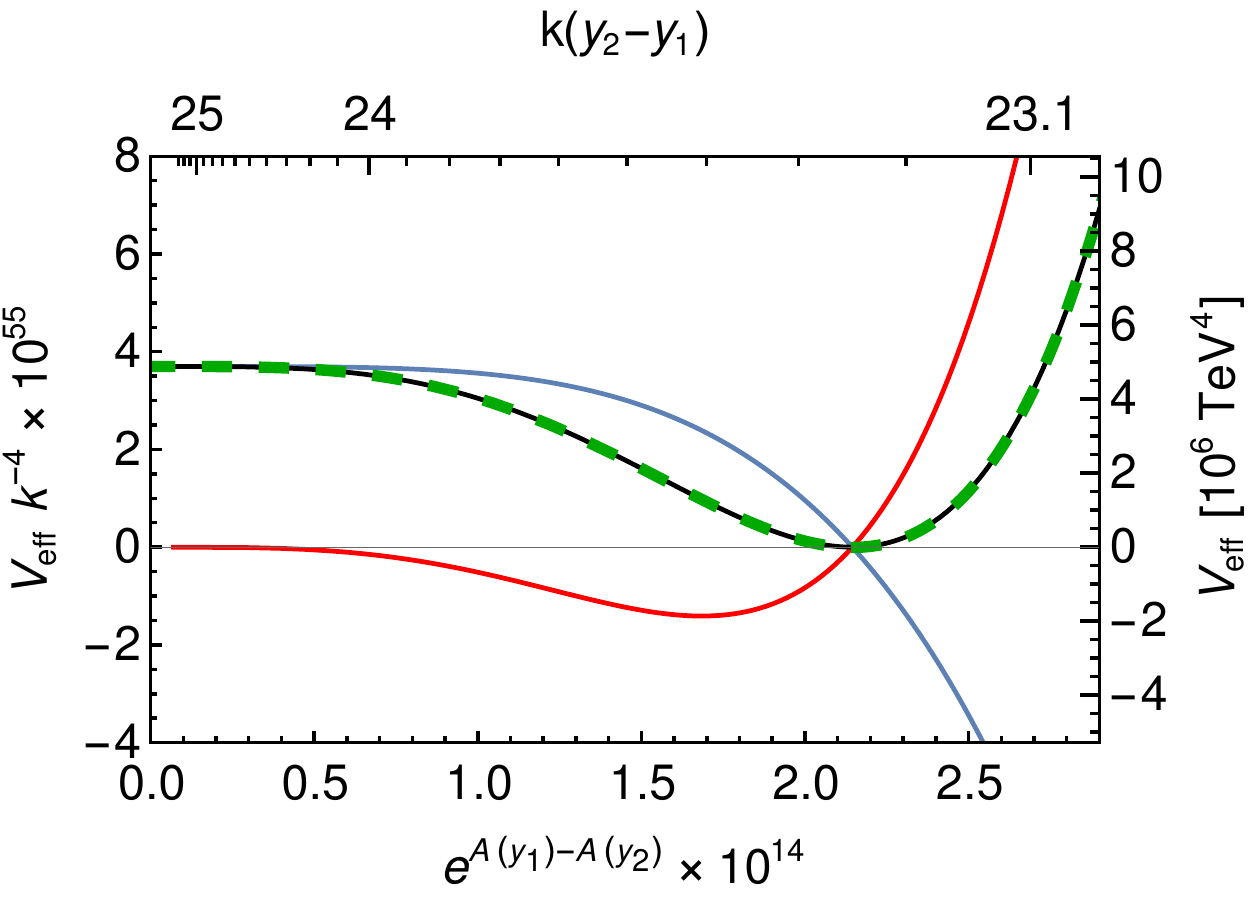}
\caption{Effective potential $V_{\mathrm{eff}}$ and its contributions as function of $\chi_g=e^{A(y_1)-A(y_2)}$ (on the upper axis we show units related to the distance between the branes:~$k\chi_y=k(y_2-y_1)$). 
The left vertical axis is given in units of $k^4$ and the right one, in TeV$^4$.
The blue and red lines are the UV ($V_{\mathrm{eff}}^{\mathrm{UV}}$), and IR ($V_{\mathrm{eff}}^{\mathrm{IR}}$) parts respectively. The black line is the total effective potential. These three curves have been computed using~\eqref{EfPot1}. The dashed green line is also the effective potential but calculated with~\eqref{VIRdV}. The agreement is perfect. On the left we show the calculation for a model with a quadratic bulk potential~\eqref{QuadBulkV}, and on the right, a model with an exponential bulk potential~\eqref{ExpBulkV}. In both cases, $\kappa=0.5\,k^{-3/2}$ and we work in the stiff wall approximation for the brane potentials. The parameters on the left are $\epsilon=0.1$, $v_1=0.106\,k^{3/2}$, $v_2=4.5\,k^{3/2}$ and $\Lambda_2=-50\,k^{4}$. On the right, $\gamma=0.1\,k^{-3/2}$, $v_1=-15\,k^{3/2}$, $v_2=-3.3\,k^{3/2}$ and $\Lambda_2=-72\,k^{4}$. The relation between $k$ and the 4D Planck mass is calculated from~\eqref{KForm}. For the left plot, $k=1.72\times10^{15}\,$TeV. For the right one, $k=1.92\times10^{15}\,$TeV.}
\label{fig:effpot1}
\end{center}
\end{figure}

In Figure~\ref{fig:effpot1} we show an example of the effective potential for both scalar potentials considered, \eqref{QuadBulkV} and \eqref{ExpBulkV}, showing at the same time the IR and UV parts defined in \eqref{VeffUV}--\eqref{VeffIR}. As one can see, both contributions are comparable in the region of the minimum, and therefore both are necessary to compute accurately the effective potential. Although the IR part has qualitatively the same shape as the full potential, the UV contribution is necessary to move the minimum of the potential to the point where $V^{\mathrm{IR}}_{\mathrm{eff}}=0$, as it is required by one of the equations of motion:~the IR Israel condition~\eqref{BC_A}\footnote{
We assume here that the configuration we consider is stable, i.e.~the distance between the branes is stabilized, so the value of the radion corresponding to this configuration is at the minimum of the effective potential.}. This shows the importance of $V^{\mathrm{UV}}_{\mathrm{eff}}$ in this region.
Under the assumptions for the parameters of the potential we consider, there is no case or limit where the UV part can be neglected, or just contributes a constant to the full potential.\footnote{Only when $\epsilon=0$, and the potential $V_{\mathrm{eff}}(\chi)\propto \chi^4+O(\chi^6)$ and the minimum of the potential is in $\chi=0$, the UV part of the potential will be a constant plus negligible corrections. This case however is not relevant to construct phenomenologically viable models.}

It has been suggested that tuning the UV brane parameters to send it to infinity (which, from the 4D point of view corresponds to send the effective 4D $M_P$ to infinity) would make $V^{\mathrm{UV}}_{\mathrm{eff}}$ to vanish. In the light of the previous argument, it cannot be the case. In the Appendix~\ref{sec:MPtoInfty} we present a calculation where we send carefully the UV brane to infinity to find the surviving contribution to the potential in such limit, and we check that $V^{\mathrm{UV}}_{\mathrm{eff}}$ does not vanish.

It is obvious from equations~\eqref{VeffUV} and \eqref{VeffIR} that, due to strong warping, $V^{\mathrm{IR}}_{\mathrm{eff}}$ is exponentially suppressed with respect to $V^{\mathrm{UV}}_{\mathrm{eff}}$. This suppression must by of around 50 or more orders of magnitude in phenomenologically realistic models. On the other hand, the two terms contributing to $V^{\mathrm{UV}}_{\mathrm{eff}}$ are $O(1)$. Therefore, if $V^{\mathrm{UV}}_{\mathrm{eff}}$ and $V^{\mathrm{IR}}_{\mathrm{eff}}$ are similar in the region of interest, $V^{\mathrm{UV}}_{\mathrm{eff}}$ must be the result of a large cancellation between the two terms in $V^{\mathrm{UV}}_{\mathrm{eff}}$. Indeed, this makes the UV part of the potential much more difficult to compute numerically than the IR part:~one has to solve the differential equations of motion keeping a very large number of significant digits (more than 50 as we have argued). This can be done, but increases the computational time. The solution to this technical difficulty is to use~\eqref{VIRdV} to compute the full potential only from the IR breaking of the boundary conditions. 
Formula~\eqref{VIRdV} allows us to find the derivative of the effective potential so an integration is necessary. As a result we obtain the potential up to an arbitrary integration constant. This constant has to be chosen in such a way that value of the potential vanishes at the minimum so that the effective 4D cosmological constant is zero. This choice corresponds to fine-tune the UV brane tension $\Lambda_1$. In all cases studied along the paper, $\Lambda_1$ will be the fine-tuned parameter to obtain a vanishing 4D cosmological constant.
In Figure~\ref{fig:effpot1} we can see the perfect agreement between the two calculation methods explained above. We emphasize that the calculation with~\eqref{EfPot1} has required to solve differential equations with more than 70 digits of significance, significantly increasing the computational time for this method. However, for the calculation using~\eqref{VIRdV}, only a few digits of significance are needed. 
This is a central result of this paper:~this new method associated with~\eqref{VIRdV} offers a clear improvement in the numeric calculation of the effective potential.

\subsection{Different choices for the interpolating radion field}

In Subsection~\ref{sec:DerivativeExp} we have discussed the dependence of the effective potential on the choice of the interpolating fields. Here we illustrate it with some examples.

As we have argued in Subsection~\ref{sec:DerivativeExp}, to define  unambiguously the potential, we should go to the Einstein frame. However, for our choice of the graviton interpolating field~\eqref{defh1} the corresponding transformations result in corrections \eqref{KeffUVmetric} which are subleading and typically very small so  we will neglect them. 
The only remaining ambiguity comes  from the choice of the radion interpolating field.
As discussed earlier, the question is how one can judge if the choice of the interpolating field is good or bad, in the  sense of giving the effective potential {\it after canonical normalization} close to the one obtained for mass eigenstates.  Since we do not calculate the coefficients of the kinetic terms, we  mimic their effects
by expressing all different effective potentials as function of the same 4D field, independently of the interpolating radion field used.
This is equivalent to applying a field redefinition after integrating out the KK modes, after which, for adequate interpolating fields we expect the kinetic terms to be similar  and, therefore, the effective potentials to look similar,
up to corrections suppressed by the integrated out heavy masses.

We will analyze the effective potential for the interpolating radion fields $\hat\chi_y$ and $\hat\chi_{\phi}$ (see Subsection~\ref{sec:OtherInterFilds}). Computing $V_{\mathrm{eff}}^{(y)}$ and $V_{\mathrm{eff}}^{(\phi)}$ we encounter similar technical features and problems as for interpolating field $\hat\chi_{g}$ discussed in the previous subsection. Although it seems more straightforward to use~\eqref{EfPot2}, it has the same numerical difficulties as~\eqref{EfPot1} for the same reasons:~we need to solve the equations of motion with the accuracy of more than 50 significant digits. It is much more economical to use~\eqref{dVeffPot2} and~\eqref{dVeffPot3}, formulae that depend only on quantities evaluated at the IR brane, and for which we need to solve the equations of motion with the accuracy of only a few significant digits.

\begin{figure}[t!]
\begin{center}
\includegraphics[height=6.5cm]{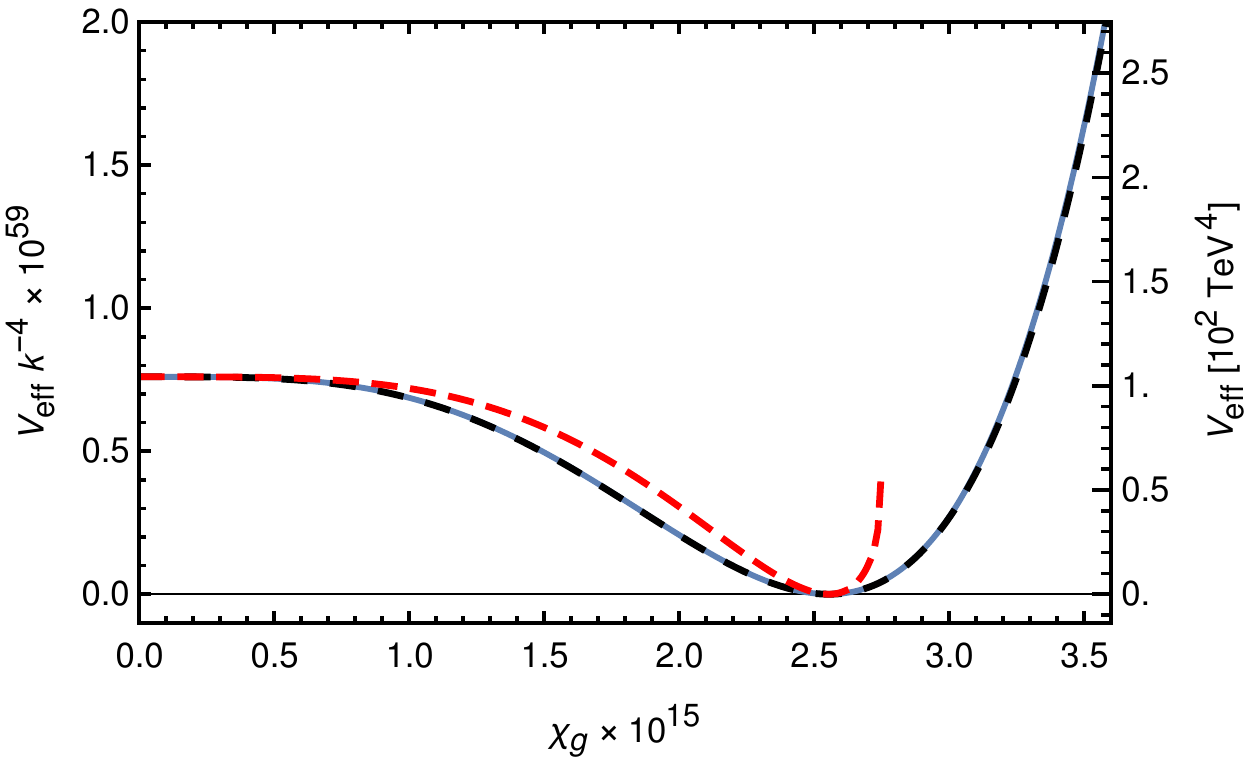}
\caption{Effective potential as a function of $\chi_g=e^{A(y_1)-A(y_2)}$. The left vertical axis is given in units of $k^4$ and the right one, in TeV$^4$. The blue line is the effective potential $V_{\mathrm{eff}}$ calculated using as interpolating radion $\hat\chi_g$. The dashed black and red lines show the effective potentials $V^{(y)}_{\mathrm{eff}}$
and $V^{(\phi)}_{\mathrm{eff}}$ respectively. 
We have used the exponential potential~\eqref{ExpBulkV} with $\kappa=0.5\,k^{-3/2}$, $\gamma=0.1\,k^{-3/2}$, $v_1=-14\,k^{3/2}$, $v_2=0.5\,k^{3/2}$, $\Lambda_2=-50\,k^{4}$, $\xi_1\to \infty$ and $\xi_2=k$.
Then, $k=1.92\times 10^{15}\,$TeV~\eqref{KForm}.}
\label{fig:Comparison1}
\end{center}
\end{figure}

\begin{figure}[t!]
\begin{center}
\includegraphics[height=4.5cm]{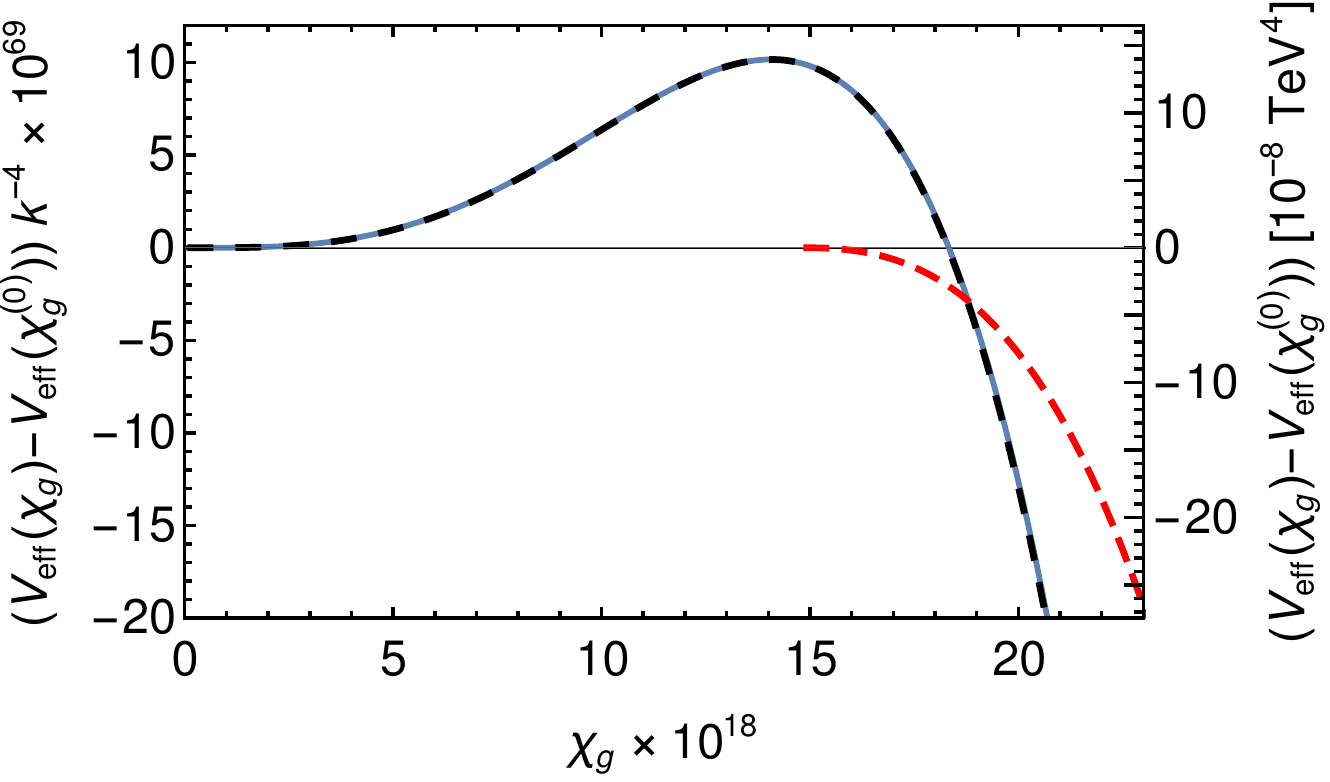}
\includegraphics[height=4.5cm]{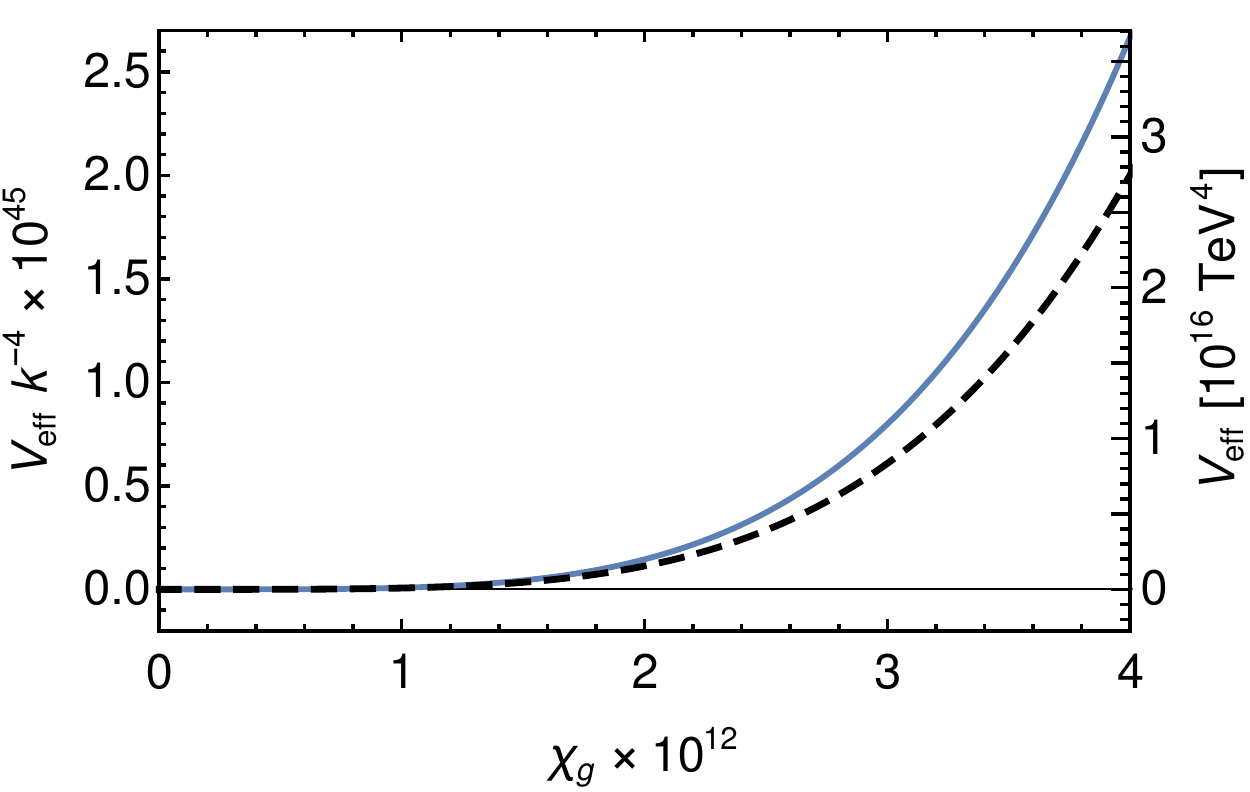}
\caption{Effective potential as a function of $\chi_g=e^{A(y_1)-A(y_2)}$. On the left the effective potentials are normalized to vanish in the lowest reachable value for $\chi_g$, $\chi_g^{(0)}$. The blue line is the effective potential $V_{\mathrm{eff}}$ calculated using as interpolating radion $\hat\chi_g$. The dashed black and red lines show the effective potentials $V^{(y)}_{\mathrm{eff}}$
and $V^{(\phi)}_{\mathrm{eff}}$ respectively.
On the left, $\chi_g$ takes values $\sim 10^{-17}$, and on the right, $\chi_g\sim 10^{-12}$.
We have used the same bulk potential as in Figure~\ref{fig:Comparison1}:~the exponential potential~\eqref{ExpBulkV} with $\kappa=0.5\,k^{-3/2}$, $\gamma=0.1\,k^{-3/2}$, $v_1=-14\,k^{3/2}$, $v_2=0.5\,k^{3/2}$, $\Lambda_2=-50\,k^{4}$, $\xi_1\to \infty$ and $\xi_2=k$.
Then, $k=1.92\times 10^{15}\,$TeV~\eqref{KForm}.}
\label{fig:Comparison2}
\end{center}
\end{figure}

To check the level of agreement between different potentials, we plot in Figures~\ref{fig:Comparison1} and~\ref{fig:Comparison2} the effective potentials $V_{\mathrm{eff}}^{(y)}$ and $V_{\mathrm{eff}}^{(\phi)}$ together with $V_{\mathrm{eff}}$ computed with the radion interpolating field $\hat \chi_g$.
We express all potentials as functions of $\chi_g=e^{A(y_1)-A(y_2)}$.

In Figure~\ref{fig:Comparison1} one can see how well the potentials $V_{\mathrm{eff}}$ and $V_{\mathrm{eff}}^{(y)}$ close to their minima agree.
This is a clear indication that both interpolating fields are adequate choices and that the assumption of the radion being very light as compared to other massive KK states is  justified.
However, $V_{\mathrm{eff}}^{(\phi)}$ does not agree with the other two. This is a consequence of a poor choice of the interpolating radion field. We have checked numerically that, in general, the contribution of the zero scalar mode to $\hat \chi_{\phi}$ is suppressed with respect the scalar KK mode contributions, so $\hat \chi_{\phi}$ is an ``almost orthogonal'' direction to the zero mode.
Also, $\hat \chi_{\phi}$ cannot be used to deform the system making $\chi_g$ arbitrary large or small because the appropriate system of EOMs has
no solutions for such values of $\chi_g$ (this is illustrated in Figures~\ref{fig:Comparison1} and~\ref{fig:Comparison2} where curves describing $V_{\mathrm eff}^{(\phi)}$ do not extend over the whole range of the radion field $\chi_g$). We anticipate that, once the canonical normalization is properly calculated for this interpolating field, the corresponding effective potential will still strongly disagree with the potential for the radion mass eigenstate.

In Figure~\ref{fig:Comparison2} we show the same potentials for two different ranges of $\chi_g$. On the left panel one can see
that both $V_{\mathrm{eff}}$ and $V_{\mathrm{eff}}^{(y)}$ agree very well reproducing the small barrier close to $\chi_g=0$, but $V_{\mathrm{eff}}^{(\phi)}$ fails and ends up at a value $\chi_g>0$.
However, differences between $V_{\mathrm{eff}}$ and $V_{\mathrm{eff}}^{(y)}$ appear for bigger ranges of the values of the radion.
We can see them on the right panel of Figure~\ref{fig:Comparison2}.
We should conclude that these disagreements indicate that this low energy description of the problem based only on the radion and the graviton and the first terms of the expansion of the effective action breaks down.
One should consider finer descriptions if such large values of the radion are to be considered.


\section{Effective potential in some models with strong back-reaction}
\label{sec:Approximate-strong-br}

As it was stated already earlier, our approach confirms the prescription for calculating the radion effective potential given by eq.~\eqref{EfPot1}, corresponding to the radion interpolating field given by the warp factor (more precisely, by the ratio of the warp factors at both branes). This prescription  has been often used in the literature, however with some approximations in the actual calculations made, especially in case of strong back-reaction. We review some of them and compare with the exact results (for the same interpolating field).

\subsection{Asymptotic matching}

If the parameter $\epsilon$ in the quadratic potential \eqref{QuadBulkV} is small one can use the method of asymptotic matching to obtain the solution to the EOMs.
This method, used in~\cite{Bellazzini:2013fga}, consists in finding analytical solutions to the equations of motion asymptotically in the deep UV and IR regions. Then, a consistency condition at some intermediate region is used to match the integration constants of both asymptotic solutions. As a result one obtains a full approximate solution. In the stiff wall limit, such approximate background solution is given by
\begin{align}
&\phi(y)\approx v_0e^{\epsilon k (y-y_1)}-\frac{\sqrt{3}}{2\kappa}\log \left[ \frac{2}{1+e^{-4k(y_2-y)}\tanh\left(\frac{\kappa}{\sqrt{3}}\left(v_2-v_1 e^{\epsilon k(y_2-y_1)} \right)  \right)}-1 \right],\\
&e^{-(A(y)-A(y_1))} \approx e^{-k(y-y_1)}\left[1-e^{-8k(y_2-y)}\tanh^2\left( \frac{\kappa}{\sqrt{3}}\left( v_2-v_1 e^{\epsilon k(y_2-y_1)}  \right) \right)  \right]^{\frac{1}{4}}.
\end{align}
These analytic expressions allow to obtain the IR part \eqref{VeffIR} of the effective potential \eqref{EfPot1} and the result reads
\begin{equation}
V_{\mathrm{eff}}^{\mathrm{IR}}\approx \frac{1}{2} \chi_g^4
\left[ \Lambda_2+\frac{6k}{\kappa^2}\cosh\left( \frac{2\kappa}{\sqrt{3}}(v_2-v_1 \chi_y^{-\epsilon}) \right) \right].\label{VIRHubAp}
\end{equation}
This expression mixes two different definition of the radion:~$\chi_g$ related to the chosen interpolating field and $\chi_y = e^{-k(y_2-y_1)}$. The latter should be expressed in terms of $\chi_g$ by inverting the relation
\begin{equation}
\chi_g=\,\chi_y \,\mathrm{sech}^{\frac{1}{2}}\left[\frac{\kappa}{\sqrt{3}}\left(v_2-v_1 \chi_y^{-\epsilon}\right)\right].\label{distForm}
\end{equation}
Such IR potential vanishes, and therefore, minimizes the full effective potential, according to~\eqref{VIRdV}, at
\begin{eqnarray}
\chi_{y,\min}&=&\left(\frac{v_1}{v_2-\mathrm{sign}(\epsilon)\frac{\sqrt{3}}{2\kappa}\mathrm{arcsech}\left(\frac{-6k}{\kappa^2\Lambda_2}\right)}\right)^{1/\epsilon},\\
\chi_{g,\min}&=&2^{\frac{1}{4}}\,\left( 1-\frac{\kappa^2\Lambda_2}{6k}\right)^{-\frac{1}{4}}\left(\frac{v_1}{v_2-\mathrm{sign}(\epsilon)\frac{\sqrt{3}}{2\kappa}\mathrm{arcsech}\left(\frac{-6k}{\kappa^2\Lambda_2}\right)}\right)^{1/\epsilon}.
\label{minAppH}
\end{eqnarray}

The above approximate solution, obtained using the method proposed in  \cite{Bellazzini:2013fga}, is not precise enough to calculate the   $V^{\mathrm{UV}}_{\mathrm{eff}}$ with accuracy sufficient to get a good approximation of the full effective potential \eqref{EfPot1} (as we argued before, many significant digits are required in the calculation of $V^{\mathrm{UV}}_{\mathrm{eff}}$). Below we propose two improvements of this asymptotic matching method. 
First, instead the usual definition of the effective potential \eqref{EfPot1}, one may apply our proposal to calculate the derivative of the effective potential solely from the IR part. Thus, using \eqref{VIRdV}, we may integrate \eqref{VIRHubAp} to obtain the full effective potential. Unfortunately, this integration cannot be done analytically. However, for small $\epsilon$, we see that $V_{\mathrm{eff}}^{\mathrm{IR}}$ in~\eqref{VIRHubAp} has a form:
\begin{equation}
V_{\mathrm{eff}}^{\mathrm{IR}}(\chi_g)= F_{\mathrm{IR}}(\chi_g)\chi_g^4,
\end{equation}
where derivatives of $F_{\mathrm{IR}}$ are small:~$\chi_g\partial_{\chi_g}F_{\mathrm{IR}}(\chi_g)={\cal{O}}(\epsilon)$ and $(\chi_g\partial_{\chi_g})^2F_{\mathrm{IR}}(\chi_g)={\cal{O}}(\epsilon^2)$.
We define analogous decomposition of the full effective potential (in this case we have to subtract its value at $\chi_g=0$ to make it vanishing at that point)
\begin{equation}
V_{\mathrm{eff}}(\chi_g)-V_{\mathrm{eff}}(0)=F(\chi_g)\chi_g^4\,.\label{FormVeff} 
\end{equation}
Using~\eqref{VIRdV}, we see that the functions $F(\chi_g)$ (for $V_{\mathrm{eff}}$) and $F_{\mathrm{IR}}(\chi_g)$ (for $V^{\mathrm{IR}}_{\mathrm{eff}}$) are related by
\begin{equation}
F(\chi_g)+\frac{1}{4}\chi_g\partial_{\chi_g}F(\chi_g)=F_{\mathrm{IR}}(\chi_g).
\end{equation}
This differential equation can be solved perturbatively in $\epsilon$ giving
\begin{equation}
F(\chi_g)=F_{\mathrm{IR}}(\chi_g)-\frac{1}{4}\chi_g\partial_{\chi_g}F_{\mathrm{IR}}(\chi_g)+{\cal{O}}(\epsilon^2).
\end{equation}
From~\eqref{VIRHubAp} we obtain
\begin{eqnarray}
F_{\mathrm{IR}} (\chi_g)&=&
\frac{1}{2}
\left[ \Lambda_2+\frac{6k}{\kappa^2}\cosh\left( \frac{2\kappa}{\sqrt{3}}(v_2-v_1 \chi_y^{-\epsilon}) \right) \right],\label{lambdaH}\\
\chi_g\partial_{\chi_g}F_{\mathrm{IR}} (\chi_g)&=&\chi_g\partial_{\chi_g}F (\chi_g)+{\cal{O}}(\epsilon^2)\nn
&=& \epsilon\,\frac{2\sqrt{3}k v_1}{\kappa}\chi_y^{-\epsilon}\,\mathrm{sinh} \left(  \frac{2\kappa}{\sqrt{3}}(v_2-v_1\chi_y^{-\epsilon} ) \right)+{\cal{O}}(\epsilon^2),\label{lambdaderivativeH}
\end{eqnarray}
so the full effective potential may be approximated as
\begin{align}
V_{\mathrm{eff}}(\chi_g)-V_{\mathrm{eff}}(0)=&\,\frac{1}{2}\, \chi_g^4
\left[ \Lambda_2+\frac{6k}{\kappa^2}\cosh\left( \frac{2\kappa}{\sqrt{3}}(v_2-v_1 \chi_y^{-\epsilon}) \right)\right.\nn
&\left.-\,\epsilon\frac{\sqrt{3}k v_1}{\kappa}\chi_y^{-\epsilon}\,\mathrm{sinh} \left(  \frac{2\kappa}{\sqrt{3}}(v_2-v_1\chi_y^{-\epsilon} ) \right)\right]+{\cal{O}}(\epsilon^2),\label{VeffPotImpAp}
\end{align}
where again $\chi_y$ should be replaced with $\chi_g$ with the help of \eqref{distForm}. 
The total effective potential $V_{\mathrm{eff}}$ differs from $V^{\mathrm{IR}}_{\mathrm{eff}}$ only by terms of order ${\cal{O}}(\epsilon)$. However, such terms are crucial to correctly locate the minimum of the potential (see~Figure~\ref{fig:effpot1}). The inclusion of these terms in the approximate expression of the potential is a new result of this article.
The depth of the potential is
\begin{align}
V_{\mathrm{eff}}(\chi_{g,\min})-V_{\mathrm{eff}}(0)&=-\frac{1}{4}\left.\chi_g\partial_{\chi_g}F_{\mathrm{IR}}(\chi_g) \chi_g^4 \right|_{\chi_{g,\min}}\nn
&=-|\epsilon| \frac{\sqrt{3}\,k}{2\kappa}\sqrt{\frac{\kappa^4\Lambda_2^2}{36 k^2}-1}
\left(v_2-\mathrm{sign}(\epsilon)\frac{\sqrt{3}}{2\kappa}\mathrm{arcsech}\left(\frac{-6k}{\kappa^2\Lambda_2}\right)\right)\chi_{g,\min}^4.\label{MinPotH}
\end{align}
It follows from~\eqref{VIRHubAp} that, in order to have a bounded from below effective potential, the parameters have to satisfy the condition
\begin{equation}
|\Lambda_2|<\frac{6k}{\kappa^2} \cosh\left[ \frac{2\kappa}{\sqrt{3}}(v_2-v_1) \right].
\label{Lambda2bound}
\end{equation}
Otherwise, $V_{\mathrm{eff}}^{\mathrm{IR}}$, and therefore, $\d V_{\mathrm{eff}}/\d \chi_g$, would be negative when $\chi_g \to \infty$.

\begin{figure}[t!]
\begin{center}
\includegraphics[height=5.7cm]{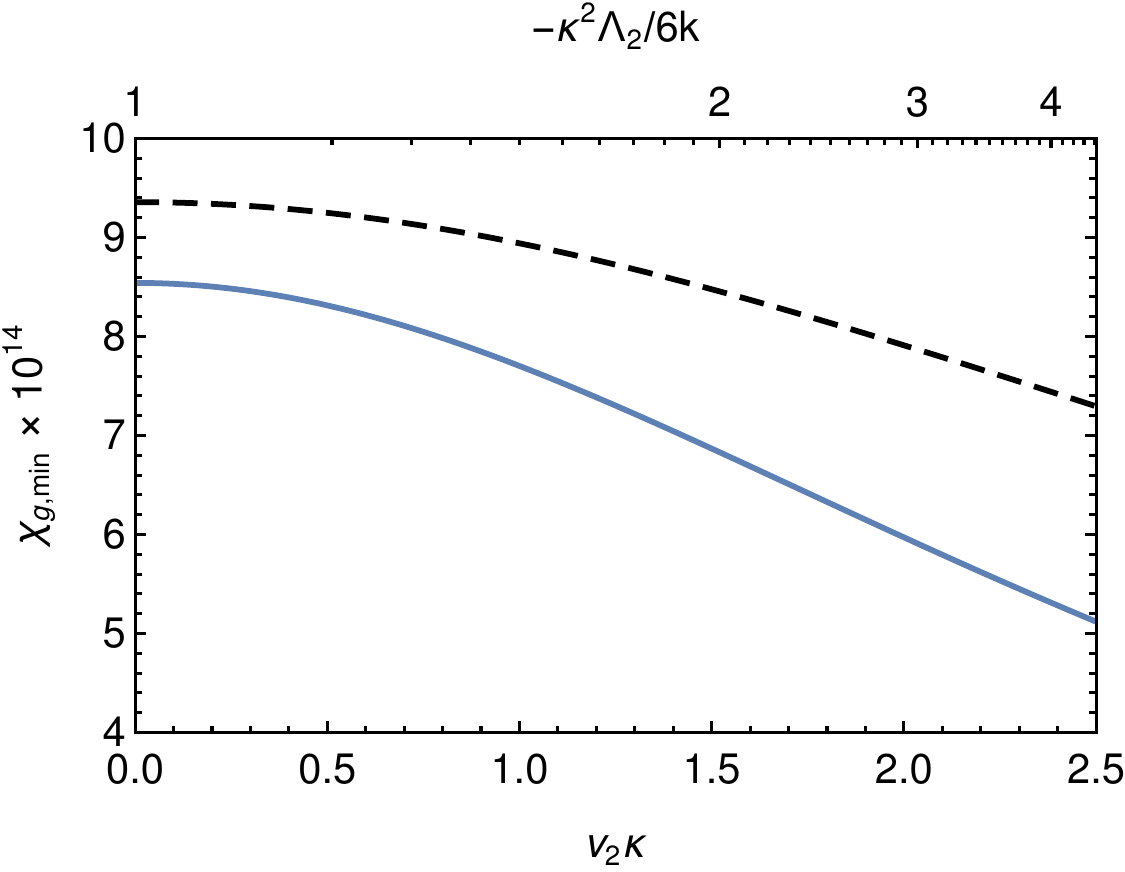}
\includegraphics[height=5.7cm]{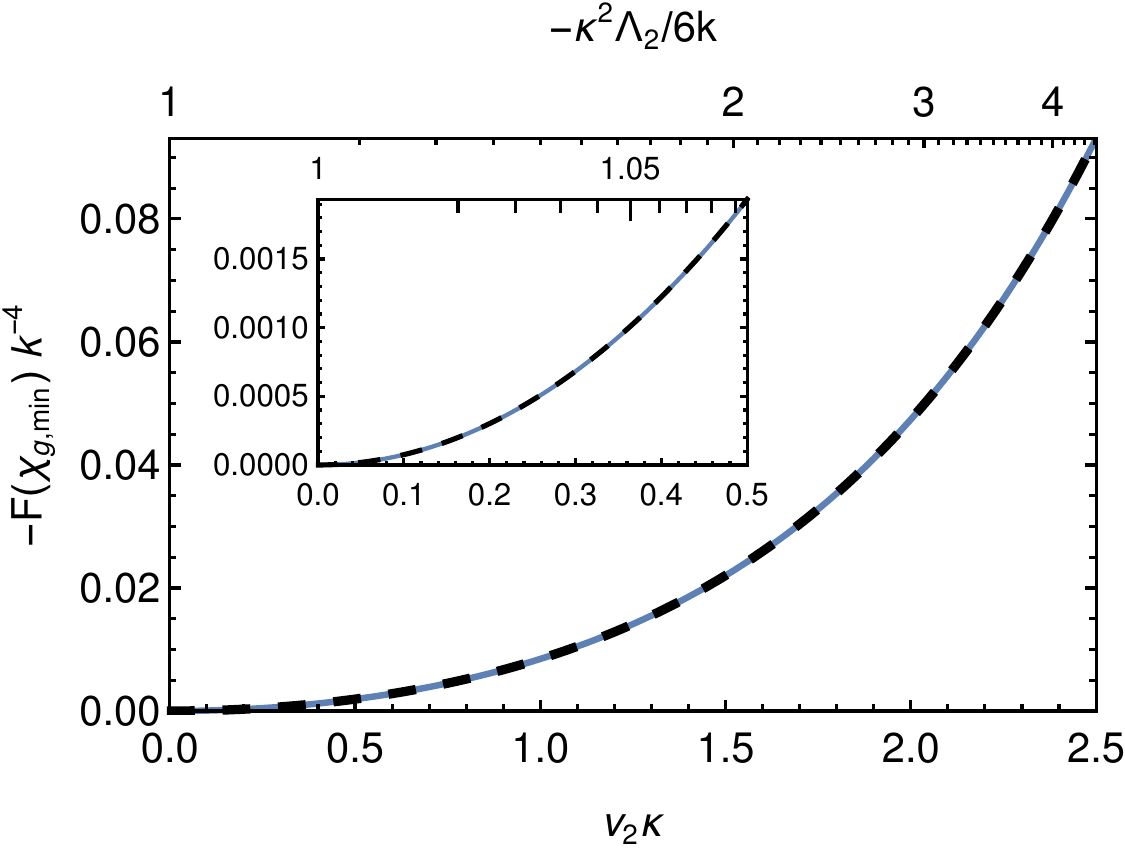}
\caption{Asymptotic matching approximation.
The choice of parameters is $\kappa=0.5\,k^{-3/2}$, $\epsilon=0.01$, $\Lambda_2=-\frac{3k}{\kappa^2}\cosh(\sqrt{3}\kappa v_2/2)$ and $v_1=(e^{-30\epsilon}/4) v_2$.
Then, $k=1.72\times 10^{15}\,$TeV~\eqref{KForm}.
On the left, the value of $\chi_{g,\mathrm{min}}$, and on the right,
 $-F(\chi_{g,\mathrm{min}})=(V_{\mathrm{eff}}(0)-V_{\mathrm{eff}}(\chi_{g,\mathrm{min}}))/\chi_{g,\mathrm{min}}^4$, both as function of $v_2\kappa$ and $-\kappa^2\Lambda_2/6k$. The blue line is the exact calculation. The black dashed line are values obtained with~\eqref{minAppH} and~\eqref{MinPotH}.}
\label{fig:MinVH}
\end{center}
\end{figure}

Now we use the benchmark model defined by
\begin{align}
\Lambda_2=&-\frac{6k}{\kappa^2}\cosh\frac{\sqrt{3}\kappa v_2}{2},\label{lambda2v2}\\
v_1=&\frac{e^{-30\epsilon}}{4}v_2\,.
\label{v1fv2}
\end{align}
In \eqref{lambda2v2} we use the r.h.s.~of~\eqref{Lambda2bound} changing $v_2 \to (3/4)v_2$ and $v_1\to 0$ in order not to saturate the bound 
but to stay close to it and make $|\Lambda_2|$ large.
The choice of~\eqref{v1fv2} comes from imposing $k(y_2-y_1)=30$ in~\eqref{minAppH}.
We use this model to judge the asymptotic matching approximation, with and without our improvement.
In Figure~\ref{fig:MinVH} we plot the value of $\chi_{g,\min}$ (left panel) and $(V_{\mathrm{eff}}(0)-V_{\mathrm{eff}}(\chi_{g,\min}))/\chi_{g,\min}^4$ (right panel) calculated exactly with the methods explained in the previous sections and within this approximation. 
On the left panel one can see that the equation~\eqref{VIRHubAp} for computing $\chi_{g,\min}$ gives inaccurate results, but correct in the order of magnitude for the whole range we have explored (which includes very large deviation from AdS geometry when $|\Lambda_2|\kappa^2/6k> 2$). On the right panel of Figure~\ref{fig:MinVH} we however see that the value obtained with the approximation for $(V_{\mathrm{eff}}(0)-V_{\mathrm{eff}}(\chi_{g,\min}))/\chi_{g,\min}^4$ matches very well with the exact result.

\begin{figure}[t!]
\begin{center}
\includegraphics[height=4.8cm]{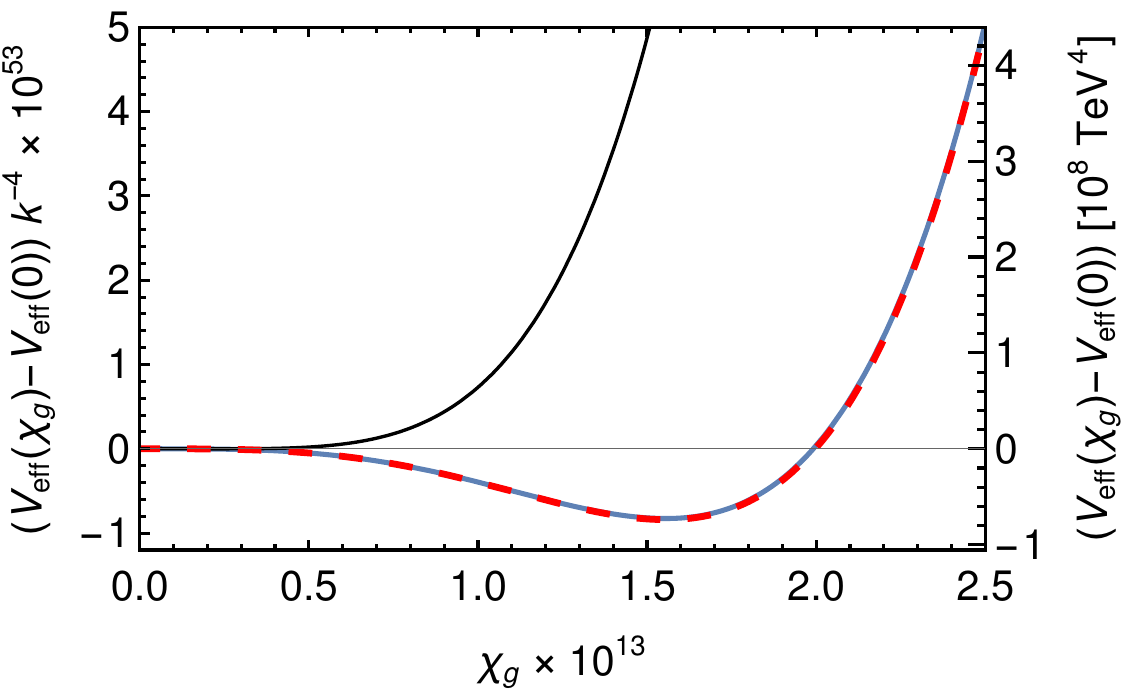}
\includegraphics[height=4.8cm]{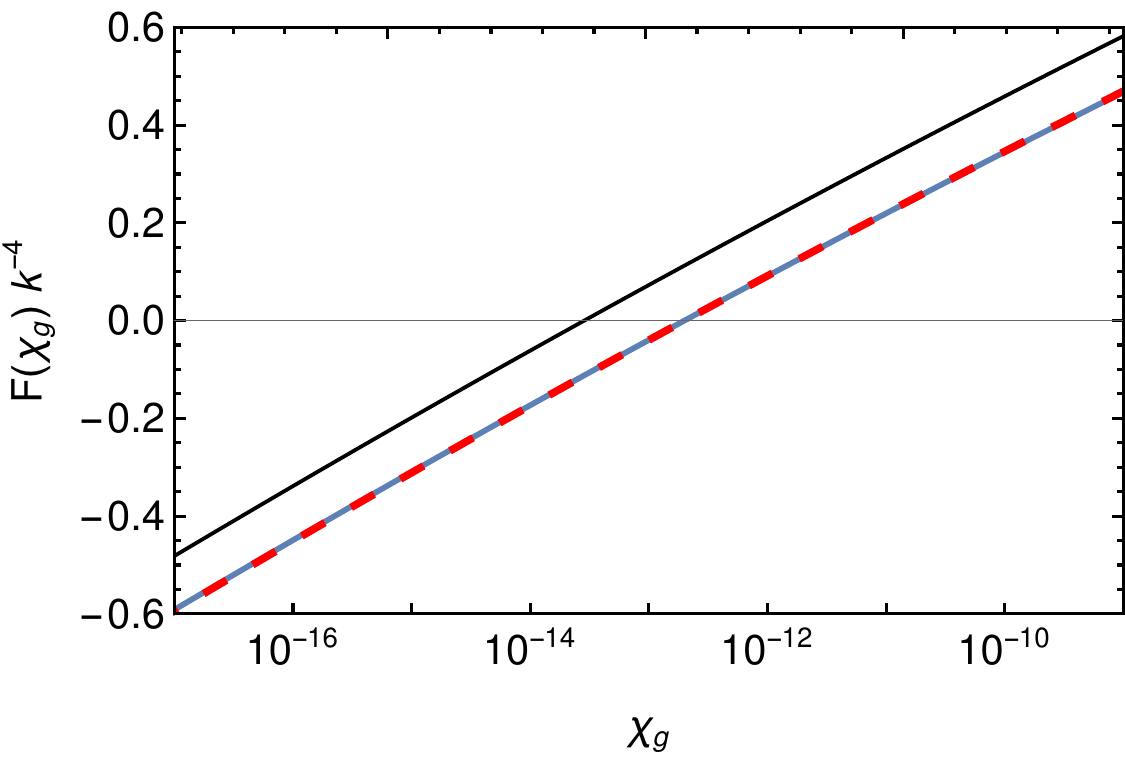}
\caption{
Exact and approximate effective potential for a system with $\kappa=0.5\,k^{-3/2}$, $\epsilon=0.01$, $v_1=0.1\,k^{3/2}$, $v_2=4.5\,k^{3/2}$ and $\Lambda_2=-150\,k^{4}$. 
Then, $k=1.72\times 10^{15}\,$TeV~\eqref{KForm}.
The black line is $V_{\mathrm{eff}}(\chi_g)-V_{\mathrm{eff}}(0)$ calculated with the approximation~\eqref{VeffPotImpAp}. The blue line gives the exact potential $V_{\mathrm{eff}}(\chi_g)-V_{\mathrm{eff}}(0)$ calculated numerically. The dashed red line (indistinguishable from the blue one) is the hybrid approximation:~\eqref{VeffPotImpAp} with the substitution~\eqref{substLBR} after computing numerically $\chi_{y,\min}$. On the left we plot the effective potential using linear scales for the axes.
The left vertical axis is given in units of $k^4$ and the right one, in TeV$^4$.
On the right, we plot $F(\chi_g)=(V_{\mathrm{eff}}(\chi_g)-V_{\mathrm{eff}}(0))/\chi_g^4$ using a logarithmic scale for the horizontal axis.}
\label{fig:EffPotH}
\end{center}
\end{figure}

This suggests that this approximation reproduces very well the quantity $\chi_g\partial_{\chi_g}F(\chi_g)$, but not the absolute value of $F(\chi_g)$, which is good enough just to reproduce the order of magnitude of $\chi_{g,\min}$. To improve these results we propose a new hybrid analytic-numeric method. 
Notice that the exact value of $\chi_{g,\min}$ is relatively easy to compute numerically: one has only to solve the equation of motion~\eqref{EoM_phi}--\eqref{EoM_A} with the correct boundary conditions~\eqref{BC_phi}--\eqref{BC_A} and evaluate $\chi_g$.
This motivates a way to improve the usual approximation. The prefactor $F(\chi_g)$ in the effective potential~\eqref{FormVeff} may be shifted by a constant in such a way that $\chi_{g,\min}$ computed in the approximation is moved to match the numerically computed value.
Therefore, we can improve these results, computing numerically the value of $\chi_{y,\min}$ before applying the approximate formulae for the effective potential. Then we apply a shift in $F(\chi_g)$ to match the value of $\chi_{y,\min}$. This is equivalent to replacing the IR brane tension in~\eqref{VeffPotImpAp} according to
\begin{equation}
\Lambda_2 \to -\frac{6k}{\kappa^2}\cosh\left( \frac{2\kappa}{\sqrt{3}}(v_2-v_1 \chi_{y,\min}^{-\epsilon}) \right).\label{substLBR}
\end{equation}
In Figure~\ref{fig:EffPotH} we plot the exact effective potential and several approximations for a particular set of parameters.
We show that the application of~\eqref{VeffPotImpAp} gives results of the same order of magnitude, but inaccurate. However, the application of the hybrid method reproduces very well the exact result.
On the right panel of Figure~\ref{fig:EffPotH} we also check that the non-improved approximation indeed estimates correctly $\chi_g\partial_{\chi_g} F(\chi_g)$~\eqref{lambdaderivativeH}, but produces a shifted $F(\chi_g)$.

\subsection{Superpotential techniques}

For the exponential potential~\eqref{ExpBulkV}, for which we know analytically the superpotential,~\cite{Megias:2018sxv} offers an alternative way to compute the effective potential for the radion if $\hat{h}_{\mu\nu}$ and $\hat \chi_g$, defined in~\eqref{defh1} and~\eqref{defchi1}, are used as interpolating fields for the graviton and the radion, respectively. The superpotential $W$ is related to the bulk scalar potential $V$ by the following differential equation
\begin{equation}
V(\phi)=\frac{1}{8}\left( \frac{\partial}{\partial \phi}W(\phi)\right)^2-\frac{\kappa^2}{6}W^2(\phi).\label{superpot}
\end{equation}
Then, functions $\phi(y)$ and $A(y)$ with derivatives given by 
\begin{equation}
\phi^{\prime}(y)=\frac{1}{2}\frac{\partial }{\partial \phi}W(\phi(y)),
\qquad
A^{\prime}(y)=\frac{\kappa^2}{6}W(\phi(y)),\label{EqWSupP}
\end{equation}
solve the bulk equations of motion~\eqref{EoM_phi}--\eqref{EoM_extra}. If in addition the superpotential satisfies the boundary conditions
\begin{eqnarray}
\frac{\partial}{\partial \phi}W(\phi(y_{1,2}))&=&\pm \frac{\partial}{\partial \phi}U_{1,2}(\phi(y_{1,2})),\label{BC_phiW}\\
W(\phi(y_{1,2}))&=&\pm U_{1,2}(\phi(y_{1,2})),\label{BC_AW}
\end{eqnarray}
the boundary conditions~\eqref{BC_phi}--\eqref{BC_A} for $\phi(y)$ and $A(y)$ are also fulfilled.

In order to get the radion effective potential one has to know bulk solutions with some of boundary conditions violated. Thus, in the case of superpotential method one has to know the whole, depending on one integration constant, family of superpotentials solving differential equation~\eqref{superpot}. In general it is impossible to find analytically all such solutions. Instead, it was proposed in~\cite{Megias:2018sxv} to expand the unknown superpotentials in some (small) parameter $s$ and to solve the equation~\eqref{superpot} perturbatively using this expansion around the one known analytic solution:
\begin{equation}
W(\phi)=W_0(\phi)+s W_1(\phi) + s^2 W_2(\phi)+ \dots\,.\label{Wexpansions}
\end{equation}
Then, the corresponding solutions of~\eqref{EqWSupP} may also be expanded in $s$:
\begin{align}
\phi(y)&=\phi_0+s\phi_1(y)+s^2\phi_2(y)+\ldots\,,\label{phiexpansions}\\
A(y)&=A_0+sA_1(y)+s^2A_2(y)+\ldots\,.\label{Aexpansions}
\end{align}
Thus, one obtaines a system of iterative differential equations for $W_1$, $W_2$, $\phi_1$, $\phi_2$, etc. The explicit formulae can be found in~\cite{Megias:2018sxv}.\footnote{In every equation for $W_n$, a new integration constant appears. They can be fixed arbitrarily because a change of them is translated into a reparametrization of the parameter $s$. We will use the convention $W_n(v_1)=0$ for $n> 1$.}
Working in the stiff wall limit~~\eqref{StiffBC_phi}--\eqref{StiffU} (this limit is not necessary for this method, but simplifies the following discussion), after breaking the boundary conditions~\eqref{BC_AW}, one finds the following expression for the effective potential
\begin{equation}
V_{\mathrm{eff}}(\chi_g)=\frac{1}{2}\left(\Lambda_1-W(v_1)\right)+\frac{1}{2}\chi_g^4(\Lambda_2+W(v_2))\,.\label{SWAV}
\end{equation}
The value of the expansion parameter $s$ and two integration constants which appear in the solutions of equations~\eqref{EqWSupP} are fixed by the considered value of the radion field and the boundary conditions~\eqref{StiffBC_phi} (conditions~\eqref{StiffBC_A} are broken):
\begin{eqnarray}
&e^{-(A(y_2)-A(y_1))}=\chi_g\,,&
\\
&\phi(y_1)=v_1\,,\qquad\phi(y_2)=v_2\,.&\label{SPBC_phi}
\end{eqnarray}
Notice that the value of $s$ does not depend on $\Lambda_2$. Therefore, $\Lambda_2$ enters the calculation only through its explicit presence in the effective potential formula~\eqref{SWAV}.

Let us now discuss this method in more detail using the specific bulk potential
~\eqref{ExpBulkV} proposed in~\cite{Megias:2018sxv}. This form of the potential follows from the analytic superpotential, which can be chosen to be $W_0$:
\begin{equation}
W_0(\phi)=\frac{6k}{\kappa^2}(1+e^{\gamma \phi})\,.\label{superpot0}
\end{equation}
For this superpotential and the IR boundary condition~\eqref{SPBC_phi}, the solutions $\phi_0$ and $A_0$ to~\eqref{EqWSupP} are
\begin{align}
\phi_0(y)=&\,v_1-\frac{1}{\gamma}\log\left(1-\frac{y-y_1}{y_s-y_1}\right),\\
A_0(y)=&\,k(y-y_1)-\frac{\kappa^2}{3\gamma^2}\log\left(1-\frac{y-y_1}{y_s-y_1}\right),\\
y_s=&\,\frac{\kappa^2}{3\gamma^2k}\mathrm{e}^{-\gamma v_1}.
\end{align}
Knowing the exact form of $W(\phi)$, i.e.~knowing all orders in $s$ (if the convergence radius of the expansion is big enough), one would get exact solutions for $\phi(y)$ and $A(y)$,~\eqref{phiexpansions} and~\eqref{Aexpansions}. However, in practice this expansion is truncated at some (rather low) order. Therefore, only the solution when $s=0$ gives an exact solution. This solution corresponds to a specific value of the interpolating radion field, which we denote by $\chi_{g,0}=\exp(A(y_1)-A(y_2))$, and in general it does not correspond to the minimum of the effective potential.
At this point ($\chi_{g,0}$), the method reproduces the exact value of the potential. It is characterized by
\begin{eqnarray}
\phi_0(y_2)=v_2 &\quad\Rightarrow\quad& y_2-y_1=\frac{\kappa^2}{3k\gamma^2} \left(e^{-\gamma v_1} - e^{-\gamma v_2}\right)\nn
&\quad\Rightarrow\quad& \log(\chi_{g,0})=-\frac{\kappa^2}{3\gamma^2} \left(e^{-\gamma v_1} - e^{-\gamma v_2 } + \gamma(v_2 - v_1)\right).
\end{eqnarray}
Assuming that $s$ is a regular function of $\chi_g$:
\begin{equation}
s(\chi_g)=s^{(1)}(\chi_{g,0}) (\chi_g-\chi_{g,0})+\frac{1}{2} s^{(2)}(\chi_{g,0})(\chi_g-\chi_{g,0})^2+\frac{1}{3!}s^{(3)} (\chi_g-\chi_{g,0})^3+\ldots
\end{equation}
we see that the truncation in~\eqref{Wexpansions} to order ${\cal{O}}(s^n)$ eliminates terms ${\cal{O}}(\chi_g-\chi_{g,0})^{n+1}$ in~\eqref{SWAV}.
Thus, the truncated version of~\eqref{SWAV} will reproduce the effective potential in the neighborhood of $\chi_{g,0}$ only up to order ${\cal{O}}(\chi_g-\chi_{g,0})^{n}$.

For physical applications, one should be able to reproduce the effective potential to a good accuracy at least in the neighborhood of its minimum. But, as we argued above, the superpotential method gives good accuracy for $\chi_g$ close to $\chi_{g,0}$ and not necessarily close to $\chi_{g,\min}$. Notice that $\chi_{g,0}$ does not depend on $\Lambda_2$, while $\chi_{g,\min}$ does. Therefore, 
the condition for $\chi_{g,\min}$ to coincide with $\chi_{g,0}$ can be achieved by  tuning the IR brane tension to be
\begin{equation}
\Lambda_2=-\frac{6k}{\kappa^2}(1+e^{\gamma v_2}) \equiv \Lambda_2^*\,.\label{lambdastar}
\end{equation}

\begin{figure}[t!]
\begin{center}
\includegraphics[height=4.7cm]{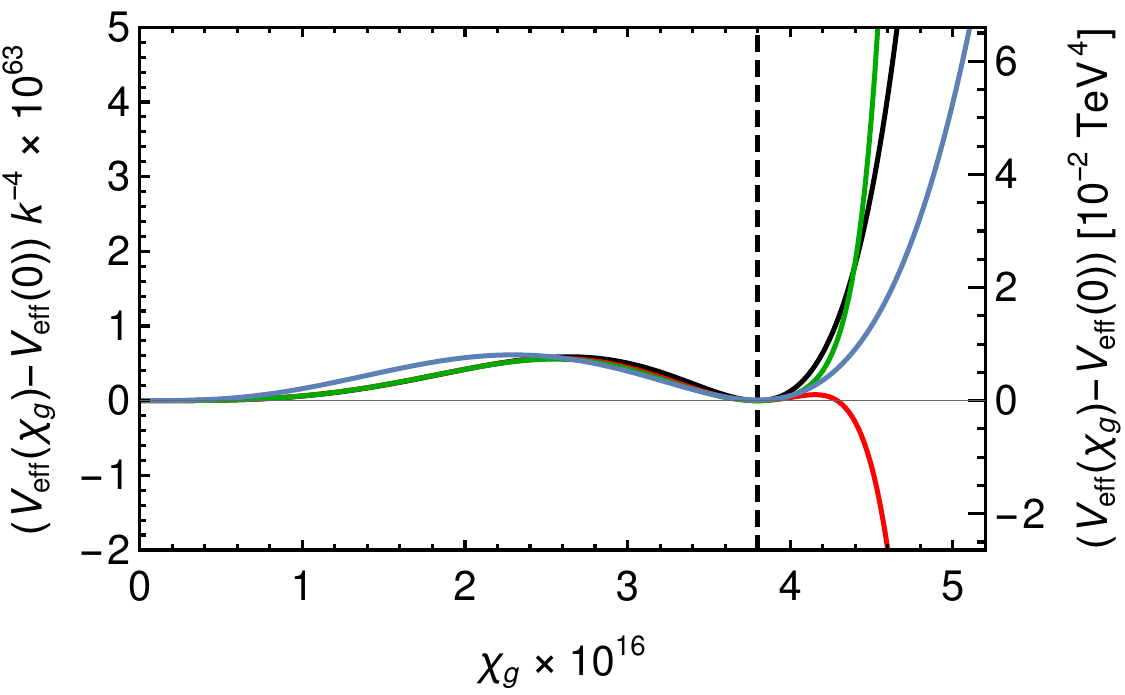}
\includegraphics[height=4.7cm]{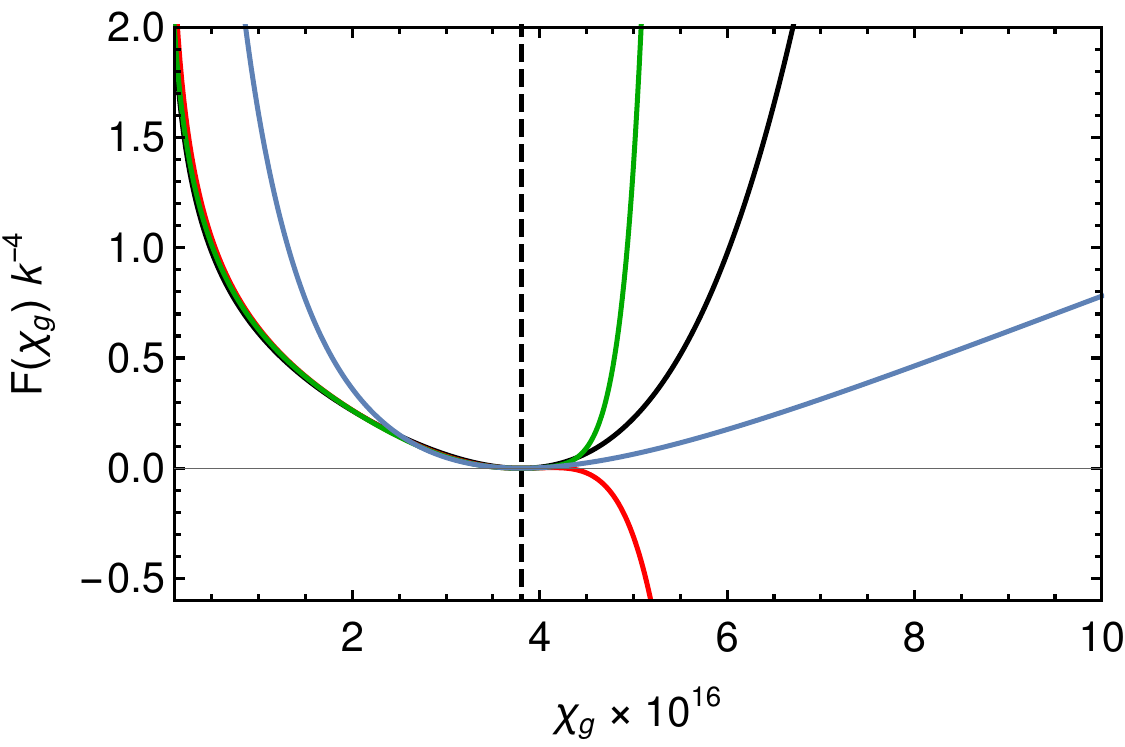}
\caption{Exact and approximate effective potential. The fixed parameters are $\kappa=0.5\,k^{-3/2}$, $\gamma=0.1\,k^{-3/2}$, $v_1=-15\,k^{3/2}$, $v_2=-3.3\,k^{3/2}$, and $\Lambda_2=\Lambda^*_2=-1.719\, \times 6k/\kappa^2$.
Then, $k=1.91\times 10^{15}\,$TeV~\eqref{KForm}.
The blue line is the exact calculation of $V_{\mathrm{eff}}$. The black, red and green lines are the approximation up to the first, second and third order respectively. 
The vertical dashed line is the value $\chi_{g,0}=3.8\times 10^{-16}$ where the approximation is exact. On the left, we plot $V_{\mathrm{eff}}(\chi_g)-V_{\mathrm{eff}}(0)$.
The left vertical axis is given in units of $k^4$ and the right one, in TeV$^4$.
On the right, we plot $ F(\chi_g)=(V_{\mathrm{eff}}(\chi_g)-V_{\mathrm{eff}}(0))/\chi_g^4$.}
\label{fig:EffPotQLS}
\end{center}
\end{figure}

To compare results of exact calculation with those obtained using the lowest orders of expansion in $s$ in the superpotential method, we take now a particular case where the condition~\eqref{lambdastar} is satisfied. In the left panel of Figure~\ref{fig:EffPotQLS} we plot the exact effective potential and the one obtained with the approximation up to order 1, 2 and 3 in the expansion in $s$. We can see that indeed the approximations reproduce the exact  effective potential in a region around $\chi_{g,0}=\chi_{g,\min}$. 
We also see that including higher orders does not improve substantially the approximation:
the second order destabilizes the potential and the third order slightly improves the first one close to $\chi_{g,0}=\chi_{g,\min}$, but then it diverges faster.
On the right panel of Figure~\ref{fig:EffPotQLS} we plot $F(\chi_g)\equiv(V_{\mathrm{eff}}(\chi_g)-V_{\mathrm{eff}}(0))/\chi_g^4$. One can see that higher orders in $s$ worsen the approximation of $F(\chi_g)$ for $\chi_g>\chi_{g,0}$ and do not improve it for $\chi_g<\chi_{g,0}$ (except a small region close to $\chi_{g,0}$).

\begin{figure}[t!]
\begin{center}
\includegraphics[height=6.5cm]{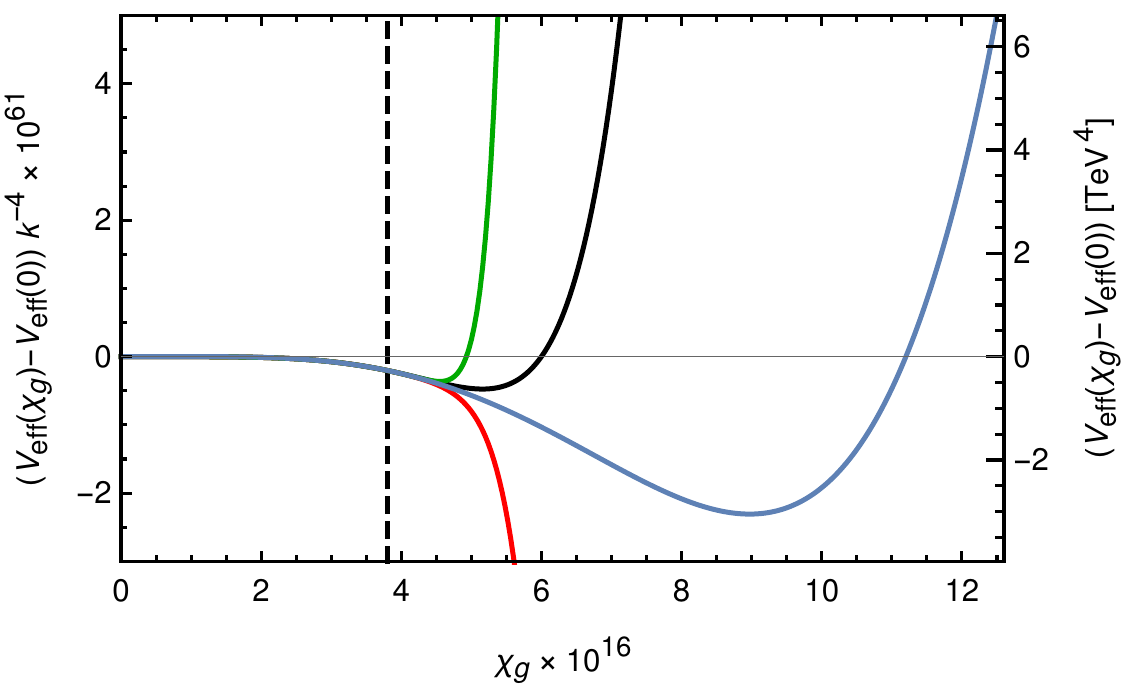}
\caption{Exact and approximate effective potential normalized to $V_{\mathrm{eff}}(0)=0$. The fixed parameters are $\kappa=0.5\,k^{-3/2}$, $\gamma=0.1\,k^{-3/2}$, $v_1=-15\,k^{3/2}$, $v_2=-3.3\,k^{3/2}$, and $\kappa^2 \Lambda_2/6k=-1.8$.
For these parameters, $\kappa^2 \Lambda^*_2/6k=-1.719$ and $k=1.91\times 10^{15}\,$TeV~\eqref{KForm}.
The left vertical axis is given in units of $k^4$ and the right one, in TeV$^4$.
The blue line is the exact calculation of $V_{\mathrm{eff}}$. The black, red and green lines are the approximation up to the first, second and third order respectively. 
The vertical dashed line is the value $\chi_{g,0}=3.8\times 10^{-16}$ where the approximation is exact.}
\label{fig:EffPotQ}
\end{center}
\end{figure}

\begin{figure}[t!]
\begin{center}
\includegraphics[height=4.5cm]{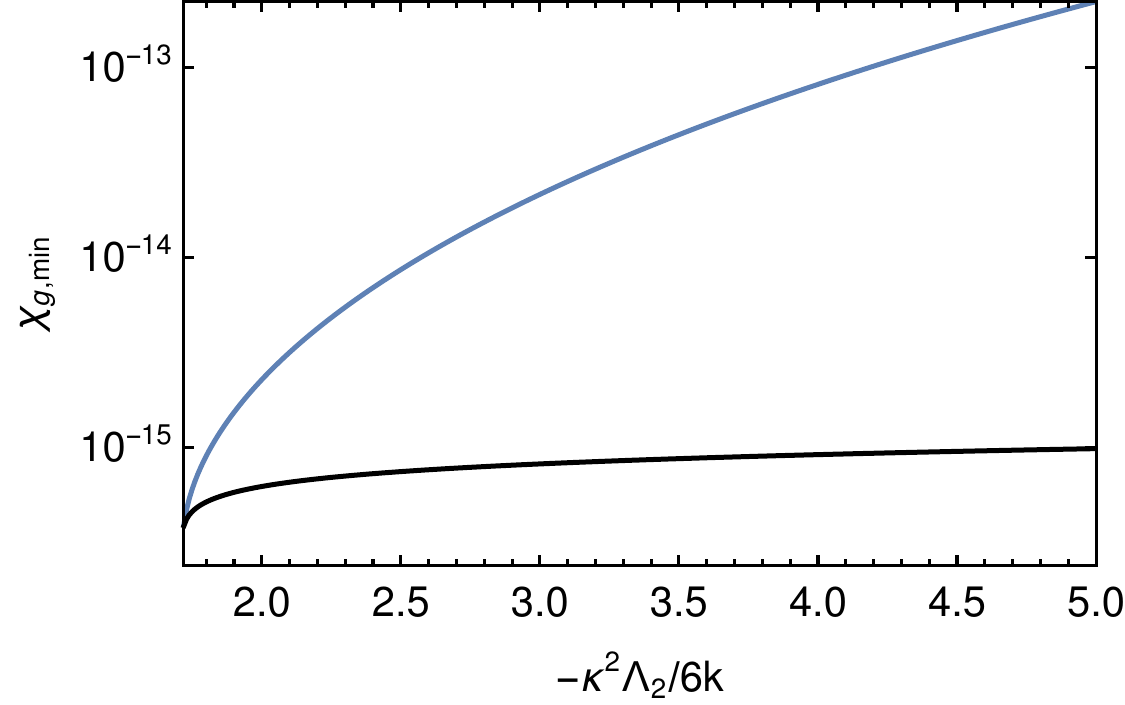}
\includegraphics[height=4.5cm]{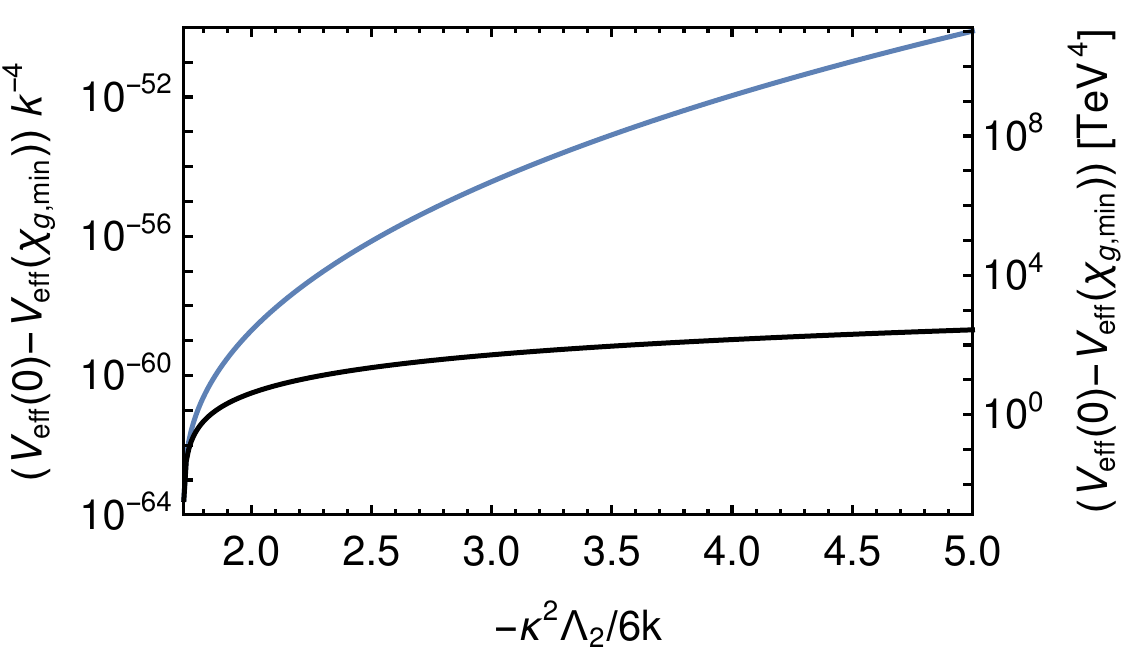}
\caption{Superpotential approximation. The choice of parameters is $\kappa=0.5\,k^{-3/2}$, $\gamma=0.1\,k^{-3/2}$, $v_1=-15\,k^{3/2}$ and $v_2=-3.3\,k^{3/2}$. For these parameters, $\kappa^2 \Lambda^*_2/6k=-1.719$ and $k=1.91\times 10^{15}\,$TeV~\eqref{KForm}. On the left, we plot the value of $\chi_{g,\mathrm{min}}$ as function of $\Lambda_2<\Lambda_2^*$. 
On the right, we plot the difference between the minimum of the potential and the origin, $V_{\mathrm{eff}}(0)-V_{\mathrm{eff}}(\chi_{g,\mathrm{min}})$ as function of $\Lambda_2<\Lambda_2^*$. In both cases, the blue line is the exact calculation and the black line the approximate calculation keeping only orders ${\cal{O}}(s)$.}
\label{fig:MinQ}
\end{center}
\end{figure}

A remarkable consequence of the choice of parameters \eqref{lambdastar} is that $V_{\mathrm{eff}}(\chi_{g,\min})-V_{\mathrm{eff}}(0)=0$.
Systems where $V_{\mathrm{eff}}(\chi_{g,\min})=V_{\mathrm{eff}}(0)$ seems to be  not interesting for constructing viable models because they cannot trigger a phase transition to a confined phase ($\chi_{g,\min}\neq 0$).
To make $V_{\mathrm{eff}}(\chi_{g,\min})<V_{\mathrm{eff}}(0)$, one has to decrease IR brane tension (see~\eqref{SWAV}), $\Lambda_2<\Lambda_2^*$. If we leave the remaining parameters unchanged, this just corresponds to a vertical shift in $F(\chi_g)$ (see the right panel of Figure~\ref{fig:EffPotQLS}). For the potential itself, it will push $\chi_{g,\min}$ to larger values $\chi_{g,\min}>\chi_{g,0}$ (while $\chi_{g,0}$ is unchanged). As may be seen in the right panel of Figure~\ref{fig:EffPotQLS}, larger values of $\chi_g~(>\chi_{g,0})$ lead soon to a poor estimation of $F(\chi_g)$, which implies a poor estimation of the effective potential in a neighborhood of the minimum (but it will still describe correctly a neighborhood of $\chi_{g,0}$).
In Figure~\ref{fig:EffPotQ} we plot the exact and approximate effective potential for a $\Lambda_2<\Lambda_2^*$. As expected, the superpotential method approximation gives a poor estimation of the potential in the region of the minimum.
To check how much the approximation disagrees with the numerically computed exact result as we decrease $\Lambda_2$, in Figure~\ref{fig:MinQ} we plot $\chi_{g,\min}$ and $V_{\mathrm{eff}}(0)-V_{\mathrm{eff}}(\chi_{g,\min})$ as functions of $\Lambda_2(<\Lambda_2^*)$ with the remaining parameters fixed as in the previous figures of this subsection. We keep only the first order in $s$ because, as discussed before, higher orders in $s$ do not improve the approximation away from $\chi_{g,0}$. Figure~\ref{fig:MinQ} shows that the results of superpotential approximate method may differ from the exact ones by, sometimes many, orders of magnitude. 

At high enough temperature, systems of this kind develop a new phase with the IR brane substituted by a black hole~\cite{Creminelli:2001th, Bunk:2017fic, Megias:2018sxv}. Above some critical temperature the free energy of such new phase is lower than that of the phase with two branes.\footnote{The critical temperature is defined as that temperature $T_c$ for which the value of the free energy for the two phases is equal.} 
Therefore, a first order phase transition takes place during the cosmological evolution of such a model. 
Neglecting finite--temperature contributions the free energy in the confined phase (phase without the black hole) is given by the corresponding effective potential. Since the difference between the value of the potential at the origin and the free energy of the black hole phase is found to be proportional to $T^4$~\cite{Creminelli:2001th,Megias:2018sxv}, the critical temperature satisfies 
\begin{equation}
T_c \propto 
\sqrt[4]{V_{\mathrm{eff}}(0)-V_{\mathrm{eff}}(\chi_{g,\mathrm{min}})}
\,.
\end{equation}
Thus, a poor estimation of the depth of the potential may result in a wrong prediction of the critical temperature. As can be inferred from Figure~\ref{fig:MinQ} such prediction obtained by approximate method in the model discussed in the present subsection may be easily off even by some orders of magnitude.


\section{Conclusions}
\label{sec:Conclusions} 

There is a continuous theoretical and experimental interest in the warped extra-dimensional  models as the framework for an extension of the Standard Model. An important ingredient of such models is the Goldberger-Wise mechanism for stabilizing the distance between the UV and IR branes. The models can address the hierarchy problem, can serve as perturbative tools to study strongly interacting scale invariant structures and may have important collider and cosmological implications. A useful concept for investigating the low energy four-dimensional effective theory  describing the gravitational and the scalar sector in the limit of heavy (decoupled) Kaluza-Klein modes is the radion effective potential.

In this technical paper, we have reviewed the concept of the radion effective potential and its subtleties. First, we have given a rigorous definition of the radion effective potential by using the interpolating field method.  
For a given choice of the interpolating field, defined as a functional of 5D fields, the radion effective action is uniquely defined by the procedure of integrating out the other fields, with the constrained 5D equations of
motion always satisfied with help of the Lagrange multipliers. Thus, for a given choice of the
interpolating field we obtain a precise prescription for calculating the effective potential in terms of the 5D parameters. Different choices of the interpolating fields give different prescriptions.  We confirm the correctness of one prescription used so far on a more
heuristic basis and also find several new, much more economical, ways of calculating the
radion effective potential.
In particular, using the Lagrange multiplier method we have found that the derivative of the effective potential with respect to the radion field is exactly given in terms of the values of the metric and the scalar field evaluated at one point in the 5-th dimension. For the case of the interpolating fields leading to the prescription 
most often used in the literature, the derivative of the effective potential depends only on quantities evaluated at the IR brane.

Once the prescription is chosen, in the next step one has to perform the actual calculations. With our new prescription, one avoids numerically very difficult  procedure of solving equations of motion with extremely high precision, to control a delicate cancellations among different contributions, especially different terms at the UV brane, in the standard approach. In the latter,  this problem is bypassed by using some approximate methods  whose quality is not easy to quantify. Having numerically easy method to calculate the exact effective potential, we have investigated its dependence on the definition of the radion interpolating field, that is on the chosen prescription. A weak dependence is an a posteriori useful check of both, the choice of the interpolating fields and the underlying assumption about the mass gap  between the zero and KK modes, underlying the effective field theory approach.

In the second part of the paper we have used our exact results for the radion effective potential calculated for various choices of the 5D parameters, to judge the quality of some approximations used in the literature with the standard prescription based on eq.~\eqref{EfPot1}.
Several new observations have been made concerning the dependence of the quality of those approximate methods on the parameter ranges of the 5D theory  and as the function of the background values of the radion field in $V_{\mathrm{eff}}(\chi)$. It is shown that in some cases those approximations, especially in models with strong
back-reaction, give results which are off even by orders of magnitude. Thus, our results are
important e.g.~for estimation of critical temperature in phase transitions.

In this paper we have not included the canonical normalization of the radion field, which requires an independent calculation of the next term in the effective action. This is of course necessary to get the final shape of the effective potential but even non-canonically normalized  it is physically interesting as it gives the correct pattern of extrema. Thus, for instance, one can discuss the character of the radion phase transition once the temperature effects are included. Those topics will be discussed in the forthcoming publication~\cite{LOP:KT}.

The tools developed in this paper can be extended in a number of ways. An effective action for more fields besides the radion and graviton can be defined. The additional fields could be the Higgs boson, to study models with Higgs-radion mixing, or even higher scalar KK modes. The last extension would open the possibility to analyze the goodness of the approximation that considers the radion as the only relevant field for the phase transition mentioned above. The improvements due to considering extra degrees of freedom can be quantified.
Also, these methods can be applied to a broader class of models that have not been considered here. 
For example models that replace the IR brane with a naked singularity
which have been also studied because of their interesting phenomenological features~\cite{Falkowski:2008fz,Cabrer:2009we,Megias:2014iwa}.

\acknowledgments
We thank M. P\'erez-Victoria, K. Sakurai and J. Serra for fruitful discussions. This work has been supported by the Polish National Science Center under the Beethoven series grant number DEC-2016/23/G/ST2/04301. MO acknowledges partial support from National Science Centre,
Poland, grant DEC-2018/31/B/ST2/02283.

\appendix

\section{Variations of the action}
\label{sec:AcVariation}

It will be useful to explicitly write the action variations~\eqref{EoM1} and~\eqref{EoM2} evaluated in the ansatz~\eqref{ansatz}--\eqref{g4}. They are given by
\begin{align}
\left.\frac{\delta S}{\delta \Phi ({\bf x},y)}\right|_{\mathrm{ansatz}}&=\,\frac{e^{-4A(y)}}{2}\sqrt{-g^{(4)}({\bf x})}\left[\phi^{\prime \prime}(y)-4\phi^{\prime}(y)A^{\prime}(y)-\frac{\partial V(\phi(y))}{\partial \phi}\right.\nn
&\hspace{6cm}\left.-\sum_{i=1,2} \frac{\partial U_{i}(\phi(y))}{\partial\phi}\delta(y-y_i)\right],\label{Var1}
\\
\left.\frac{\delta S}{\delta g^{\mu\nu} ({\bf x},y)}\right|_{\mathrm{ansatz}}&=\,g^{(4)}_{\mu\nu}\frac{3e^{-6A(y)}}{4\kappa^2 }\sqrt{-g^{(4)}({\bf x})}\Bigg[ 2A^{\prime 2}(y)-A^{\prime \prime}(y)+
\frac{\kappa^2}{6}\phi^{\prime 2}(y)+\frac{\kappa^2}{3}V(\phi(y)) \nn
&\hspace{3.9cm}-e^{2A(y)}H^2+\frac{\kappa^2}{3}\sum_{i=1,2} U_{i}(\phi(y))
\delta(y-y_i)\Bigg],\label{EoM3b}\\
\left.\frac{\delta S}{\delta g^{55} ({\bf x},y)}\right|_{\mathrm{ansatz}}&=\,\frac{e^{-4A(y)}}{4}\sqrt{-g^{(4)}({\bf x})} \left[\frac{6}{\kappa^2}A^{\prime 2}(y)-\frac{\phi^{\prime 2}(y)}{2} +V(\phi(y))-\frac{6}{\kappa^2}e^{2A(y)}H^2\right].\label{EoM2b}
\end{align} 
Setting them to zero one obtains~\eqref{EoM_phi}-\eqref{EoM_extra}.

\section{Ansatz method to obtain the 4D effective action}
\label{sec:AnsatzMethod} 

Let us consider a 5D theory with 5D fields that we generically call $\omega({\bf x},y)$.
In the ansatz method, used e.g.~in~\cite{Luty:2000ec, Bagger:2000eh, Bagger:2003dy, Chacko:2013dra}, the first step is to choose an ansatz for the 5D fields as functions of the light 4D fields $h_{\mu\nu}({\bf x})$, $\chi({\bf x})$:
\begin{equation}
\omega({\bf x},y)=\hat \Omega_0[h_{\mu\nu},\chi]({\bf x},y).\label{ansatzA2}
\end{equation}
Most of the choices are not consistent with the 5D EOMs. In such cases the light fields inevitably develop heavy field tadpoles, i.e.~terms in the Lagrangian linear in the heavy fields, which have to be integrated out.
To analyze this possibility, let us complete the ansatz~\eqref{ansatzA2} to a 
full parametrization of the 5D fields in terms of light fields $h_{\mu\nu}$, $\chi$ and heavy fields $\psi_i$:
\begin{equation}
\omega({\bf x},y)=\hat \Omega[h_{\mu\nu},\chi,\psi_i]({\bf x},y),\label{GenansatzA2}
\end{equation}
such that, when the heavy fields vanish, we recover~\eqref{ansatzA2},
\begin{equation}
\hat \Omega[h_{\mu\nu},\chi,0]=\hat \Omega_0[h_{\mu\nu},\chi].\label{GenansatzCond}
\end{equation}
This completion is non unique, but any well-defined one may be used. With parametrization \eqref{GenansatzA2} we can express the 5D action as a function of the light and heavy fields and obtain the corresponding 4D action
\begin{equation}
S=S_{\mathrm{light}}[h_{\mu\nu},\chi] +\int \d {\bf x}\, \psi_i({\bf x}) \,\lambda^i[h_{\mu\nu},\chi]({\bf x}) + S_{Q}[h_{\mu\nu},\chi,\psi_i],\label{NewL}
\end{equation}
where $S_{Q}$ depends at least quadratically on $\psi_i$. The ansatz~\eqref{ansatzA2} is compatible with the 5D EOMs if there are non-trivial solutions for the EOMs derived from~\eqref{NewL} with $\psi_i=0$. This is possible if the heavy field tadpoles $\lambda^i[h_{\mu\nu},\chi]$ vanish or are proportional to the EOMs obtained from $S_{\mathrm{light}}$:
\begin{equation}
\lambda^i[h_{\mu\nu},\chi]({\bf x})=\int \d {\bf x}^{\prime}\left(J^i_{h\,\mu\nu}[h_{\mu\nu},\chi]({\bf x},{\bf x}^{\prime}) \frac{\delta S_{\mathrm{light}}}{\delta h_{\mu\nu}({\bf x}^{\prime})}+J^i_{\chi}[h_{\mu\nu},\chi]({\bf x},{\bf x}^{\prime})\frac{\delta S_{\mathrm{light}}}{\delta \chi({\bf x}^{\prime})}\right).
\end{equation}
In the latter case, the fields $h_{\mu\nu}$ and $\chi$ can be redefined is such way that~\eqref{GenansatzCond} stays satisfied and the tadpoles $\lambda^i$ become $0$:
\begin{align}
h_{\mu\nu}^{\prime}({\bf x})=&\,h_{\mu\nu}({\bf x})-\int \d {\bf x}^{\prime}\,\psi_i({\bf x}^{\prime})\, J^i_{h\,\mu\nu}[h_{\mu\nu},\chi]({\bf x}^{\prime},{\bf x}),\nn
\chi^{\prime}({\bf x})=&\,\chi({\bf x})-\int \d {\bf x}^{\prime}\,\psi_i({\bf x}^{\prime}) \,J^i_{\chi}[h_{\mu\nu},\chi]({\bf x}^{\prime},{\bf x}).\label{redefinition}
\end{align}
The 4D action obtained after this field redefinition could also be directly derived choosing an adequate ansatz completion~\eqref{GenansatzA2}. 

If the ansatz \eqref{ansatzA2} is consistent with the 5D EOMs, it is possible to choose a completion \eqref{GenansatzA2} without heavy fields tadpoles, so the tree level integration of the heavy fields does not give corrections. The 4D effective action for $h_{\mu\nu}$ and $\chi$ is given by $S_{\mathrm{light}}[h_{\mu\nu},\chi]$. If the completion~\eqref{GenansatzA2} were not properly chosen, the effective action after integrating out the heavy field tadpoles would be different than $S_{\mathrm{light}}[h_{\mu\nu},\chi]$, but related by a field redefinition.
On the other hand, if the ansatz \eqref{ansatzA2} is not consistent with the 5D EOMs, there are always heavy field tadpoles which must be integrated out.

Once the completion of the ansatz~\eqref{GenansatzA2} is chosen, we can invert it to find:
\begin{align}
h_{\mu\nu}=&\hat h_{\mu\nu}[\omega],\label{hIF}\\
\chi=&\hat \chi[\omega],\label{chiIF}\\
\psi_i=&\hat \psi_i[\omega].
\end{align}
We stress that the form of~\eqref{hIF} and~\eqref{chiIF} depends not only on the chosen ansatz~\eqref{ansatzA2} but also on how the heavy fields enter~\eqref{GenansatzA2}.
The functional form of the interpolating light fields~\eqref{hIF}--\eqref{chiIF} determines the resulting 4D effective action.

The interpolating field method used in this paper avoids the intermediate steps of obtaining~\eqref{NewL} and integrating out heavy fields tadpoles if they appear. Once the interpolating light fields~\eqref{hIF}--\eqref{chiIF} are specified, we directly integrate out the remaining degrees of freedom. The result is the same as the one obtained with the ansatz method after integrating out tadpoles, with the adequate choice of the ansatz~\eqref{ansatzA2} and its completion~\eqref{GenansatzA2}.

\section{Calculation of the kinetic mixing with gravity}
\label{sec:KineticMixingG}

To obtain the function $K_{\mathrm{eff}}$ of~\eqref{EffActV}, we have to explore solutions with non-flat 4D metric $h_{\mu\nu}$. This can be done restricting to homogeneous 4D configurations $\chi({\bf x})=\chi$ and $h_{\mu\nu}= g^{(4)}_{\mu\nu}$, where $g^{(4)}_{\mu\nu}$ is given in~\eqref{g4} with general $H$.
In this case, the effective action is
\begin{equation}
\mathcal{L}_{\mathrm{eff}}=-V_{\mathrm{eff}}+6 K_{\mathrm{eff}} H^2 + {\cal{O}}(H^4).\label{LEffHneq0}
\end{equation}
Due to symmetry considerations, given $\chi$ and $H$, the solution to the system of equations~\eqref{EoM1}--\eqref{EoM4}, $\Phi_{\mathrm{sol}}$ and $g_{\mathrm{sol}}$, can be written using the ansatz~\eqref{ansatz}--\eqref{g4} with the same value of $H$. The solution is fixed in terms of the functions $\phi_{\mathrm{sol}}(y)$ and $A_{\mathrm{sol}}(y)$.

We insert now the solution of the system in~\eqref{LEffIntegr5}. For 5D fields with the form of~\eqref{ansatz}--\eqref{g4}, the 5D Ricci scalar can be decomposed as
\begin{equation}
R^{(5)}[g]=e^{2A}\,R^{(4)}[g^{(4)}]+R^{(5)}[\bar g(A)],
\end{equation}
where $R^{(D)}[g]$ is the Ricci scalar in $D$ dimensions for the metric $g$ and $\bar g(A)$ is the 5D metric with 4D flat sections with the warp factor $e^{-2A}$:
\begin{equation}
\bar g_{MN}(A)\d x^M \d x^N=\d y^2 + e^{-2A} \eta_{\mu\nu} \d x^{\mu} \d x^{\nu}.
\end{equation}
Using this and the fact that the scalar field does not depend on ${\bf x}$, we can write~\eqref{LEffIntegr5} as
\begin{equation}
\mathcal{L}_{\mathrm{eff}}(h_{\mu\nu},\chi)=
\frac{6H^2}{\kappa^2}\int_{y_1}^{y_2}\d y\,e^{-2A_{\mathrm{sol}}}
+\frac{1 }{2}\int_{S^1}\d y \,\sqrt{\bar g(A_{\mathrm{sol}})}\, \mathcal{L}_{\mathrm{5D}}(y,\bar g(A_{\mathrm{sol}}),\Phi_{\mathrm{sol}}),\label{LEffSplit}
\end{equation}
where we have used equality $R^{(4)}[g^{(4)}]=12 H^2$.
The explicit form of the functions $\Phi_{\mathrm{sol}}$ and $A_{\mathrm{sol}}$, given $\chi$ and $H$, depends on the particular choice of the interpolating fields. In any case, they can be expanded for small $H^2$:
\begin{align}
\Phi_{\mathrm{sol}}=&\Phi_{(0)}+H^2\Phi_{(2)}+O(H^4),\\
A_{\mathrm{sol}}=&A_{(0)}+H^2A_{(2)}+O(H^4).
\end{align}
Inserting this expansion in~\eqref{LEffSplit} we have
\begin{align}
\mathcal{L}_{\mathrm{eff}}=&
\frac{1 }{2}\int_{S^1}\d y \,\sqrt{\bar g(A_{(0)})}\, \mathcal{L}_{\mathrm{5D}}(y,\bar g(A_{(0)}),\Phi_{(0)})+
\frac{6H^2}{\kappa^2}\int_{y_1}^{y_2}\d y\,e^{-2A_{\mathrm{sol}}}\nn
&+\frac{H^2}{2}\int_{S^1} \d y 
\left[\frac{\delta S[\bar g(A_{(0)}),\Phi_{(0)}]}{\delta g^{MN} }
\frac{\partial  \bar g^{MN}(A_{(0)})}{\partial A}A_{(2)}+
\frac{\delta S[\bar g(A_{(0)}),\Phi_{(0)}]}{\delta \Phi}
\Phi_{(2)}
\right]+O(H^4).\label{LIntStep}
\end{align}
The first term gives $-V_{\mathrm{eff}}$. In the second line we have expressed the variation of the Lagrangian as the variation of the action. A total derivative would appear additionally, but it vanishes under the integral in $y$. We can now use the equations of motion~\eqref{EoM1}--\eqref{EoM2} and express the variation of the action as the variation of the interpolating fields. To continue with the argumentation, we need to assume a property of the interpolating fields. Their variation with respect to 5D fields that are 4D homogeneous and 4D flat, $\Phi=\phi(y)$ and $g_{MN}=\bar g_{MN}(A)$, must be
\begin{equation}
\left.\frac{\delta \hat h_{\rho\sigma}({\bf x})}{\delta g^{\mu\nu}({\bf x}^{\prime},y)},~
\frac{\delta \hat \chi({\bf x})}{\delta g^{\mu\nu}({\bf x}^{\prime},y)}\right|_{\bar g(A),\Phi} \propto \delta({\bf x}-{\bf x}^{\prime}).\label{ConditionIntF}
\end{equation}
We will only consider interpolating fields satisfying this condition. Then,~\eqref{LIntStep} can be re-expressed as
\begin{align}
\mathcal{L}_{\mathrm{eff}}=&-V_{\mathrm{eff}}(\chi)+
\frac{6H^2}{\kappa^2}\int_{y_1}^{y_2}\d y\,e^{-2A_{\mathrm{sol}}}\nn
&+H^2\left.\int \d {\bf x}\left( \lambda^h_{\mu\nu}\frac{\partial}{\partial H^2} \hat h^{\mu\nu} [\bar g(A_{\mathrm{sol}}),\Phi_{\mathrm{sol}}]
+\lambda^{\chi}\frac{\partial}{\partial H^2} \hat \chi [\bar g(A_{\mathrm{sol}}),\Phi_{\mathrm{sol}}]
\right)\right|_{H^2=0}
+O(H^4),\label{LIntStep2}
\end{align}
where, as stated above, $A_{\mathrm{sol}}$ and $\Phi_{\mathrm{sol}}$ depends on $H^2$. Given some configuration of the 5D fields $\Phi$ and $g$ having the form of the anszat~\eqref{ansatz}--\eqref{g4} with any functions $\phi$ and $A$ (non necessarily a solution of the system), the interpolating fields can be written as $\hat h_{\mu\nu}[g,\Phi]=g^{(4)}_{\mu\nu}\, \hat f_h[A,\phi,H^2]$ and $\hat \chi[g,\Phi]=\hat f_{\chi}[A,\phi,H^2]$. Here, $\hat f_h$ and $\hat f_{\chi}$ are functionals of $A$ and $\phi$, and functions of $H^2$ (the interpolating fields could depend on the curvature tensors of $g^{(4)}$). To satisfy~\eqref{ConditionIntF}, $\hat f_{h}$ and $\hat f_{\chi}$ must be such that
\begin{equation}
\left.\frac{\partial}{\partial H^2}\hat f_{h,\chi}[A,\phi,H^2]\right|_{H^2=0}=0.
\end{equation}
On the other hand, $\hat f_{h,\chi}[A_{\mathrm{sol}},\phi_{\mathrm{sol}},H^2]$ must be independent of $H^2$, where now $A_{\mathrm{sol}}$ and $\phi_{\mathrm{sol}}$ are the solutions for a given $H^2$: the interpolating fields are fixed to be $g^{(4)}$ and $\chi$. Then, the derivatives in the second line of~\eqref{LIntStep2} vanish. Comparing~\eqref{LIntStep2} and~\eqref{LEffHneq0}, we can extract the function $K_{\mathrm{eff}}(\chi)$ describing the kinetic mixing with gravity:
\begin{equation}
K_{\mathrm{eff}}(\chi)= \frac{1}{\kappa^2}  \int_{y_1}^{y_2} \d y\, e^{-2 A_{\mathrm{sol}}}\,,\label{KFormApp}
\end{equation}
where $A_{\mathrm{sol}}$ is the solution for given homogeneous $\chi$ and $h_{\mu\nu}=\eta_{\mu\nu}$.

One can check that for models discussed in this work, and for the graviton interpolating field \eqref{defh1}, the radion-gravity mixing depends very weakly of $\chi$. Analyzing the perturbations of the equations of motion~\eqref{EoM_phi}--\eqref{EoM_extra}, we can obtain that the variation of the warp factor $A(y)\to A(y)+\delta A(y)$ for a variation of the interpolating radion field $\chi\to \chi+\delta \chi$ is order 
\begin{equation}
\delta A(y)= {\cal{O}}\left(e^{4(A(y)-A(y_2))}\right)\, \delta \chi / \chi.
\end{equation}
Such variation of $\chi$ produces a variation of $K_{\mathrm{eff}}(\chi)$ which, using~\eqref{KFormApp}, will be order
\begin{equation}
\delta K_{\mathrm{eff}}(\chi)=\frac{1}{\kappa^2}\int_{y_1}^{y_2}\d y\, {\cal{O}}\left(e^{2A(y)-4A(y_2))}\right)\, \delta \chi / \chi
={\cal{O}}\left(e^{-2(A(y_2)-2A(y_1))}\right)\,\delta \chi/\chi\,.
\end{equation}
We can conclude that, given some reference value for the radion $\chi^*$,
\begin{align}
K_{\mathrm{eff}}(\chi)
=K_{\mathrm{eff}}(\chi^*)+\frac{\chi-\chi^*}{\chi^*}\, {\cal{O}}\left(e^{-2(A(y_2)-2A(y_1))}\right)\,.
\label{KeffUVmetric}
\end{align}
Then, for $\chi$ in the range that produces large values of $A(y_2)-A(y_1)$, the change of $K_{\mathrm{eff}}(\chi)$ with $\chi$ will be suppressed, so $K_{\mathrm{eff}}(\chi)$ can be approximated to be constant. This is precisely the physically interesting range. The 4D Planck mass squared can be calculated with~\eqref{KFormApp} using the background function $A_{\mathrm{sol}}$ for some value of the  radion chosen in that range.

\section{Effective potential in the 4D $\mathbf{M_P\to\infty}$ limit}
\label{sec:MPtoInfty}

Under the AdS/CFT duality~\cite{Maldacena:1997re}, the models we investigate describe a strongly coupled conformal field theory coupled at the Planck scale to dynamical gravity and to some weakly coupled sector (also called elementary sector~\cite{Gherghetta:2010cj}).
The UV brane is responsible for this interaction and explicitly breaks the conformal symmetry through irrelevant terms~\cite{ArkaniHamed:2000ds,Rattazzi:2000hs,PerezVictoria:2001pa}. The radion is holographically dual to a dilaton, and its existence points out that the conformal symmetry is spontaneously broken around the IR scale of the model. The IR brane is actually the origin of such breaking, and it is hologrophically related to an operator of the CFT of arbitrary high dimension developing a vacuum expectation value~\cite{ArkaniHamed:2000ds,Rattazzi:2000hs}. If the distance between the branes is large, and so the hierarchy between the Planck scale and IR scale, the model may be decsribed in an approximate conformal regime below the Planck scale (explicit breaking) and above the IR scale (spontaneous breaking). A light dilaton may appear in an explicitly broken CFT without fine-tuning, as discussed in~\cite{Bellazzini:2013fga,Coradeschi:2013gda}. This can be achieved by the so-called Contino-Pomarol-Rattazzi mechanism~\cite{CPR}.

The effective potential of the radion receives contributions, not only from the strongly coupled sector, but also from the elementary sector. If the hierarchy between the scales is large, the elementary sector and gravity tend to decouple from the CFT. 
As we will see, in the limit in which the 4D Planck scale is sent to infinity, the contribution from the elementary sector reduces to a deformation of the CFT through a unique operator, the one dual to the bulk scalar field.

Sending the effective 4D Planck mass to infinity is equivalent to eliminating the UV brane by sending it to $y\to-\infty$. The resulting  4D theory does not contain dynamical gravity,
so it can live in a Minkowski space without requiring the usual fine-tuning necessary to adjust the 4D cosmological constant.
To calculate the radion effective potential in such a case we start with the ansatz~\eqref{ansatz}--\eqref{ansatz2} with $g^{(4)}_{\mu\nu}=\eta_{\mu\nu}$. The holographic coordinate $y$ runs from the IR brane at $y_2$ to $-\infty$. 
The EOMs are given by~\eqref{EoM_phi}-\eqref{EoM_extra} with $H=0$.
Let us assume that the bulk potential has an extremum, which, without loss of generality, may be chosen to be located at $\Phi=0$ where it has a negative value. The potential close to such extremum may be expanded as
\begin{equation}
V(\Phi)=-\frac{6k^2}{\kappa^2}-2\epsilon k^2\Phi^2+ {\cal{O}}(\Phi^3).
\end{equation}
For $y\to-\infty$ and $\Phi \to 0$, the solution to~\eqref{EoM_phi}-\eqref{EoM_extra} may be written in the form
\begin{align}
\phi(y)=&\,\Phi_{(1,0)}  e^{\Delta_- ky}+ \Phi_{(2,0)}  e^{2\Delta_- ky}  +\ldots+\Phi_{(0,1)}e^{\Delta_+ ky}+\Phi_{(0,2)}e^{2\Delta_+ ky}+\ldots\nn
+&\Phi_{(1,1)}e^{(\Delta_-+\Delta_+) ky}+\ldots+\Phi_{(n,m)} e^{(n\Delta_-+m\Delta_+) ky}+\ldots,~~n,m\in\mathbb{N},~n+m\geq 1,\label{AsympPhi} \\
A(y)=&\,ky+A_0+A_{(2,0)}e^{2\Delta_- ky}+A_{(3,0)}e^{3\Delta_- ky}+\ldots+A_{(0,2)}e^{2\Delta_+ ky}+A_{(0,3)}e^{3\Delta_+ ky}+\ldots\nn
+&A_{(1,1)}e^{(\Delta_-+\Delta_+) ky}+\ldots+A_{(n,m)}e^{(n\Delta_-+m\Delta_+) ky}+\ldots,~~n,m\in\mathbb{N},~n+m\geq 2,\label{AsympA}
\end{align}
where
$\Delta_{\pm}=2\pm 2 \sqrt{1-\epsilon}$.\footnote{
For 4D non-homogeneous backgrounds, each term has to be substituted by an appropriate tower of terms
\begin{align}
\Phi_{(n,m)}&\to\Phi^{(0)}_{(n,m)}+\Phi^{(2)}_{(n,m)}e^{2 k y}+\Phi^{(4)}_{(n,m)}e^{4 k y}+\ldots\\
A_{(n,m)}&\to A^{(0)}_{(n,m)}+A^{(2)}_{(n,m)}e^{2 k y}+A^{(4)}_{(n,m)}e^{4 k y}+\ldots
\end{align}
Additionally, we are assuming that every term in the sum has a different exponent. Otherwise, polynomials of $y$ may also appear in front of the exponential terms (for instance, if $n\Delta_-+m\Delta_+=n^{\prime}\Delta_-+m^{\prime}\Delta_+$ for two different pairs of $(n,m)\neq (n^{\prime},m^{\prime})$).}
All the coefficients in this expansion are determined by the EOMs if we fix $\Phi_{(1,0)}$ and $\Phi_{(0,1)}$. The coefficients $\Phi_{(n,m)}$ and $A_{(n,m)}$ depend on $\Phi_{(1,0)}$ ($\Phi_{(0,1)}$) only if $n>0$ ($m>0$). In addition, all coefficients $\Phi_{(n,m)}$ and $A_{(n,m)}$ with $n>0$ ($m>0$) vanish if $\Phi_{(1,0)}=0$ ($\Phi_{(0,1)}=0$).
If $\epsilon>0$, then $\Delta_{\pm}>0$, and $\Phi\to 0$ and $A(y)\sim ky$ when $y\to-\infty$. The space-time is therefore asymptotically AdS (AAdS) with the conformal boundary at $y\to-\infty$.
If  $\epsilon<0$, then $\Delta_-<0$ and $\Delta_+>0$, and only solutions with $\Phi_{(1,0)}=0$ have such asymptotic behavior.

Following the AdS/CFT dictionary, these models are holographically dual to a CFT deformed by $\Phi_{(1,0)}\mathcal{O}$, where $\mathcal{O}$ is a relevant (irrelevant) operator if $\epsilon>0$ ($\epsilon<0$) with dimension $\Delta_+$~\cite{Witten:1998qj,Gubser:1998bc}. The coefficient~$\Phi_{(0,1)}$ is related to the vacuum expectation value of $\mathcal{O}$~\cite{Gubser:1998bc}.\footnote{
In general, it is possible to define two different quantum theories for a scalar living in an AAdS space, using one of two possible quantization proceduress~\cite{Klebanov:1999tb}. We use here the standard quantization which may be applied for all values of $\epsilon\leq1$. With the alternative quantization, the dual theory and the dictionary change. The roles of $\Phi_{(1,0)}$ and $\Phi_{(0,1)}$ are exchanged, so $\Phi_{(0,1)}$ is related to the source and $\Phi_{(1,0)}$ to the expectation value. The dimension of the dual operator becomes equal to $\Delta_-$. However, the alternative quantization exists only for values $3/4<\epsilon<1$. Values of $\epsilon>1$ are not allowed for a field in an AAdS space due to the Breitenlohner-Freedman bound~\cite{Breitenlohner:1982jf,Klebanov:1999tb}.}
We will only consider models with $\epsilon>0$. They have a well controlled $y\to-\infty$ limit. Their dual CFTs are then deformed by a relevant operator and have a well-behaved UV limit.

The counting of the degrees of freedom of the background solution is as follows. After imposing the bulk equations~\eqref{EoM_phi}-\eqref{EoM_extra}, the solution depends on three integration constants:~$\Phi_{(1,0)}$, $\Phi_{(0,1)}$ and $A_0$ in~\eqref{AsympPhi} and~\eqref{AsympA}. Also, the position of the IR brane, $y_2$, is an extra parameter to be fixed.
The constant $A_0$ may be set to zero using the shift symmetry of the EOMs, which can be seen as a choice of units.
In the IR brane we have two boundary conditions,~\eqref{BC_phi} and~\eqref{BC_A}. These leave only one degree of freedom, and relate $\Phi_{(1,0)}$, $\Phi_{(0,1)}$ and $y_2$.
If we fix $y_2$, we can read $\Phi_{(1,0)}$ and $\Phi_{(0,1)}$ in the UV. Notice that the shift $y_2 \to y_2+a$ affects other coefficients as
\begin{align}
\Phi_{(1,0)} \to & \,  e^{-\Delta_{-} k a}\Phi_{(1,0)},\nn
\Phi_{(0,1)} \to & \, e^{-\Delta_{+} k a}\Phi_{(0,1)}. 
\end{align}
Therefore, unless there is some symmetry or fine-tuning, we would not expect a vanishing $\Phi_{(1,0)}$ for any finite $y_2$:~the dual CFT is necessarily deformed by the dual operator $\mathcal{O}$.
The model does not have any UV brane, so no UV boundary conditions~\eqref{BC_phi} and~\eqref{BC_A} have to be imposed. Instead, for a scalar field in an AAdS space, we must impose one additional boundary condition associated to the asymptotic behavior of the field in the UV~\cite{Papadimitriou:2007sj}. We take the value of $\Phi_{(1,0)}$ as a  boundary condition. This is a very natural choice because it defines the source of the deformation of the dual CFT.
Because we have changed a model with UV brane and two UV boundary conditions to a model without UV brane and one UV boundary condition, no fine-tuning is necessary to obtain solutions with flat 4D sections as we anticipated.

To compute the radion effective potential for this model, let us specify the interpolating fields to be used:
\begin{align}
\hat h_{\mu\nu}[g,\Phi]&=\lim_{y\to-\infty} e^{-2ky}\,g_{\mu\nu}(y),\label{defmetricMPI}\\
\hat \chi[g,\Phi]&=\lim_{y\to-\infty}e^{-ky} \left(-g|_y\right)^{-1/8} \left(-g|_{y_2}\right)^{1/8},\label{defchi1inf}
\end{align}
where $g|_{y}$ is the metric in the hypersurface $y$ inherited from $g$.
Using~\eqref{AsympA}, these interpolating fields are $\hat h_{\mu\nu}=e^{-2A_0}\eta_{\mu\nu}$ and $\hat \chi=\left.e^{-(A-A_0)}\right|_{y_2}$. Applying the Lagrange multiplier method, these constraints break the IR Israel junction condition~\eqref{BC_A}, which allows to change the position of the IR brane without changing $\Phi_{(1,0)}$ or breaking any other equation of motion.

The effective potential is given by minus the on shell action of homogeneous and flat configurations. However, it is well known that on shell actions in an AAdS space are divergent due to infinite volume of the AdS space. Such divergences are holographically related to the UV divergences of the dual quantum field theory~\cite{Susskind:1998dq}. A renormalization procedure to remove them and to obtain a finite (renormalized) action $S_{\mathrm{ren}}$ is necessary. There are several approaches known under the name of holographic renormalization~\cite{Bianchi:2001kw,Heemskerk:2010hk,Lizana:2015hqb}. Here, we will follow the most standard one~\cite{Skenderis:2002wp}. It consists in:~(1)~computing the regularized on shell action $S_{\mathrm{reg}}$ by integrating over $y$ to some finite value $y_1$; (2)~adding a suitable counterterm action $S_{\mathrm{c.t}}$ localized at $y_1$, and then (3)~sending $y_1\to -\infty$:
\begin{equation}
S_{\mathrm{ren}}=\lim_{y_1\to-\infty}\left(S^{y_1}_{\mathrm{reg}}[g,\Phi]+S_{\mathrm{c.t.}}[g|_{y_1},\Phi(y_1)]\right).
\end{equation}
Here, $g$ and $\Phi$ are the 5D solutions to the EOMs (breaking the IR boundary condition~\eqref{BC_A}). 
No new UV boundary conditions that affect the on shell fields $g$ and $\Phi$ have to be considered:~$g$ and $\Phi$ are kept unaltered during this procedure so they are the solution when the value $\Phi_{(1,0)}$ is fixed in the $y_1\to-\infty$ limit.
The effective potential is
\begin{align}
V_{\mathrm{eff}}=&\left.\frac{1}{2}e^{-4(A-A_0)}\left[\frac{6}{\kappa^2}A^{\prime}+U_2(\Phi)\right]\right|_{y=y_2^-}\nn
&+\lim_{y_1\to-\infty}
\left.\frac{1}{2}e^{-4(A-A_0)}\left[-\frac{6}{\kappa^2}A^{\prime}-2\mathcal{L}_{\mathrm{c.t.}}(g|_y,\Phi)\right]\right|_{y=y_1^+},\label{VeffSct}
\end{align}
where $A_0$ is the coefficient of the expansion~\eqref{AsympA} and
the couterterm action has been written as
\begin{equation}
S_{\mathrm{c.t.}}[g|_y,\Phi]=\int \d ^4x \sqrt{- g|_y}\,\mathcal{L}_{\mathrm{c.t.}}(g|_y,\Phi).
\end{equation}
Inserting the asymptotic expansions~\eqref{AsympPhi} and~\eqref{AsympA} in the second line of~\eqref{VeffSct}, we can extract the required form for $\mathcal{L}_{\mathrm{c.t.}}$ to cancel the divergences:
\begin{equation}
\mathcal{L}_{\mathrm{c.t.}}(g|_y,\Phi)=-\frac{3k}{\kappa^2}-k\frac{\Delta_{-}}{2}\Phi^2 +{\cal{O}}(\Phi^3).
\end{equation}
For non-homogeneous configurations or more general metric backgrounds, additional terms have to be introduced~\cite{deHaro:2000vlm}.
In the $y_1\to-\infty$ limit we obtain
\begin{align}
V_{\mathrm{eff}}=&\left.\frac{1}{2}e^{-4(A-A_0)}\left[\frac{6}{\kappa^2}A^{\prime}+U_2(\Phi)\right]\right|_{y=y_2^-}+\frac{k}{2}e^{4A_0}\Delta_{-}(\Delta_{-}-2)\Phi_{(1,0)}\Phi_{(0,1)}.\label{Veffminf}
\end{align}
The effective potential is therefore written as a contribution from the fields evaluated at the IR brane plus a non-vanishing contribution given by the asymptotic behavior of the fields in the UV.

\bibliographystyle{JHEP}
\bibliography{Radion}

\end{document}